\documentclass{fopstyle_unofficial}
\usepackage{mathtools,bm} 
\usepackage{cite}
\usepackage{epstopdf} 
\usepackage{array,booktabs}
\graphicspath{{figures/}}

\newcommand{\ket}[1]{\left|#1\right\rangle}
\newcommand{\bra}[1]{\left\langle#1\right|}

\newcommand{\FOPincludegraphics}[2][]{%
  \IfFileExists{figures/#2}{\includegraphics[#1]{#2}}{%
    \IfFileExists{#2}{\includegraphics[#1]{#2}}{%
      \fbox{\begin{minipage}[c][18mm][c]{0.76\linewidth}
       \centering\scriptsize Missing original figure:\par
       \texttt{\detokenize{#2}}
      \end{minipage}}%
    }%
  }%
}

\let\oldtheorem\theorem
\let\endoldtheorem\endtheorem
\renewenvironment{theorem}
{\oldtheorem\normalfont}
{\endoldtheorem}

\let\oldlemma\lemma
\let\endoldlemma\endlemma
\renewenvironment{lemma}
{\oldlemma\normalfont}
{\endoldlemma}

\FOPtitle{Local distinguishability of five five-partite orthogonal product states}
\FOPauthors{Guang-Bao Xu$^{1,2}$, Jie Zhan$^{1}$, Yu-Guang Yang$^{3}$ and Dong-Huan Jiang$^{4}$}

\FOPaffiliation{1. College of Computer Science and Engineering, Shandong University of Science and Technology, Qingdao, 266590, China\\
                2. Shandong Key Laboratory of Smart Mine Information Technology, Qingdao 266590, China\\
                3. College of Cyberspace Science and Technology, Beijing University of Technology,Beijing 100124, China\\
                4. College of Mathematics and Systems Science, Shandong University of Science and Technology, Qingdao 266590, China}
\FOPemail{Corresponding author: donghuan\_jiang@163.com}
\FOPabstract{%
	The local distinguishability of five orthogonal product states (OPSs) has been studied in several multipartite settings, but the five-partite case still requires a systematic characterization. In this paper, we consider five five-partite OPSs, where any two states are orthogonal on exactly one subsystem. By analyzing the distribution of ten pairwise orthogonality relations over five subsystems, we classify the corresponding orthogonality structures by colored orthogonality graphs. After eliminating identical graphs under relabeling of states and permutation of subsystems, we obtain $443$ different graphs in seven genuinely five-partite categories. We prove that the OPSs corresponding to 434 of these graphs can be perfectly distinguished by local operations and classical communication (LOCC), while the remaining $9$ exceptional cases fall into two types and require separate analysis. Our results give a characterization of the local distinguishability of five OPSs in five-partite OPSs.
	}
\FOPkeywords{local operations and classical communication, orthogonal product states, quantum state discrimination, quantum nonlocality}

\begin{document}
\FOPfrontmatter

\section{Introduction}
	\label{sec:introduction}
	
	In a distributed quantum system, several separated parties may share an unknown state and try to determine which state has been prepared. If a joint measurement on the whole system is allowed, mutually orthogonal quantum states can be discriminated perfectly. However, such a global operation is usually not available when the parties are restricted to local operations and classical communication (LOCC). Thus, it is necessary to determine whether a set of orthogonal states can be perfectly distinguished by LOCC when designing quantum protocols.                                                                                                                                                                                                     
	
	Bennett et al.~\cite{Bennett1999QuantumNonlocality} constructed a set of orthogonal product states (OPSs) that cannot be perfectly distinguished by LOCC, and they called this phenomenon quantum nonlocality without entanglement. Bennett et al.~\cite{Bennett1999UPB} also proved that the states forming unextendible product bases cannot be perfectly distinguished by LOCC. Further connections among unextendible product bases, uncompletable product bases, and bound entanglement were studied by DiVincenzo et al.~\cite{DiVincenzo2003Unextendible}. After these works, many basic results on LOCC discrimination were established. Walgate et al.~\cite{Walgate2000LocalDistinguishability} proved that any two orthogonal pure states can be perfectly distinguished by LOCC. Walgate and Hardy~\cite{Walgate2002Nonlocality} further investigated the local distinguishability of bipartite quantum states. DiVincenzo et al.~\cite{DiVincenzo2003Unextendible} showed that any bipartite set with at most four OPSs, and any multipartite set with at most three OPSs, is locally distinguishable.
	
	On the other hand, many studies have focused on constructing locally indistinguishable sets of OPSs. Nonlocality of orthogonal product bases and OPSs has been investigated in different bipartite and multipartite systems \cite{Zhang2014NonlocalityOPB,Zhang2015NonlocalityOPS,Wang2015NonlocalityOPB,Xu2016MultipartiteNonlocality,Xu2016ArbitraryBipartite}. Further construction methods for nonlocal sets of OPSs were developed in multipartite systems \cite{Zhang2017Construction,Halder2018SeveralNonlocalSets,Jiang2020NonlocalSetsArbitrary,Zhu2022TripartiteLessElements,Zhen2022GeneralMultipartite}.

	Other related studies have focused on different aspects of quantum nonlocality without entanglement. Halder et al.~\cite{Halder2019StrongQuantumNonlocality} introduced strong quantum nonlocality without entanglement. Strong nonlocality and related constructions were further studied in multipartite systems \cite{Yuan2020StrongQuantumNonlocality,Shi2020StrongWithEntanglement,Wang2021GraphConnectivity,Shi2022UPBHeterogeneous,Shi2022NPartite,Zhou2023EvenN}. Minimal nonlocality and locally stable nonlocal sets were investigated in Refs.~\cite{Zhu2023MinimalNonlocality,Xu2025MinimalNonlocality,Zhen2025LocallyStable}.	
	
	Recently, the local distinguishability of sets containing a fixed number of OPSs has attracted considerable attention. Xu et al.~\cite{Xu2025FourMultipartiteOPS} considered four multipartite OPSs and used the vector of the numbers of pairwise orthogonal relations to describe the distribution of orthogonality relations among subsystems. Xu et al.~\cite{Xu2026FiveBipartiteTripartiteOPS} studied the local distinguishability of five OPSs on bipartite and tripartite quantum systems. Nevertheless, the local distinguishability of five OPSs on five-partite quantum systems has remained unknow.
	
	In this paper, we consider the local distinguishability of five five-partite OPSs, where any two of the five states are orthogonal on exactly one subsystem. By analyzing the distribution of ten pairwise orthogonality relations among five subsystems, we classify the genuinely five-partite cases into seven categories and enumerate their orthogonality graphs. Up to relabeling of states and permutation of subsystems, we obtain $443$ different orthogonality graphs in these seven categories. We prove that the sets of five OPSs corresponding to 434 of these graphs can be perfectly distinguished by LOCC, while the remaining nine graphs fall into two exceptional types. The rest of this paper is organized as follows. In Sec.~\ref{sec:preliminaries}, some necessary definitions and preliminary results are introduced. In Sec.~\ref{sec:distinguishable}, we give the local distinguishability of ﬁve OPSs. In Sec.~\ref{sec:exceptional}, the 9 cases that cannot be perfectly distinguished by LOCC are discussed. A brief conclusion is given in Sec.~\ref{sec:conclusion}.

	
	\section{Preliminaries}
	\label{sec:preliminaries}
	
	In this section, some necessary definitions and preliminary results are introduced.
	
	\begin{definition}
		\label{def:orthogonality-graph}\cite{DiVincenzo2003Unextendible}
		Let $\mathcal{H}=\bigotimes^{m}_{i=1}\mathcal{H}_{i}$ be an $m$-partite Hilbert space with $\dim\mathcal{H}_{i}=d_{i}$. Let $S=\{\vert\psi_{j}\rangle\equiv \bigotimes^{m}_{i=1}\vert\varphi_{i,j}\rangle \mid j=1,2,\ldots,n\}$ be a set of mutually orthogonal product states in $\mathcal{H}$. We represent $S$ as a graph $G=(V,E_{1}\cup E_{2}\cup\cdots\cup E_{m})$, where the edges in $E_{i}$ have color $i$. The states $\vert\psi_{j}\rangle\in S$ are represented as the vertices $V$. There exists an edge $e$ of color $i$ between the vertices $v_{k}$ and $v_{l}$, i.e., $e\in E_{i}$, when the corresponding local states $\vert\varphi_{i,k}\rangle$ and $\vert\varphi_{i,l}\rangle$ are orthogonal on $\mathcal{H}_{i}$. Since all the states in $S$ are mutually orthogonal, every vertex is connected to all the other vertices by at least one edge of some color. The graph $G$ is called the orthogonality graph of $S$.
	\end{definition}

	\begin{definition}
		\label{def:orthogonality-number-vector}
		For five five-partite OPSs, a quintuple $(a,b,c,d,e)$ is called the vector of the numbers of pairwise orthogonal relations, where $a$ denotes the number of pairwise orthogonal relations on the first subsystem, while $b$, $c$, $d$, and $e$ denote the corresponding numbers on the second, third, fourth, and fifth subsystems, respectively.
	\end{definition}
	\begin{definition}
		\label{def:local-degree}\cite{BondyMurty2008GraphTheory}
		Let $G=(V,\,E_1\cup E_2\cup E_3\cup E_4\cup E_5)$ be the orthogonality graph of five five-partite OPSs, and let $v_j\in V$ be a vertex of $G$. The degree of the $i$-th party of $v_j$, denoted by $\deg_i(v_j)$, is the number of edges incident with $v_j$ that belong to $E_i$.
	\end{definition}
	\begin{definition}
		\label{def:orthogonal-neighborhood}
		Let $v$ be a vertex of the orthogonality graph $G=(V,\,E_1\cup E_2\cup E_3\cup E_4\cup E_5)$. For each subsystem $p\in\{1,2,3,4,5\}$, let $N_p(v)=\{u\in V\setminus\{v\}\mid \{u,v\}\in E_p\}$ denote the set of vertices whose corresponding states are orthogonal to the state corresponding to $v$ on the $p$-th subsystem. Clearly, $|N_p(v)|=\deg_p(v)$.
	\end{definition}
	\begin{definition}
		\label{def:local-rank}
		\cite{DiVincenzo2003Unextendible,Xu2025FourMultipartiteOPS}
		Let $\mathcal{H}=\bigotimes_{i=1}^{m}\mathcal{H}_i$ be an $m$-partite Hilbert space with $\dim\mathcal{H}_i=d_i$, and let $\mathcal{S}=\{\ket{\psi_j}\equiv\bigotimes_{i=1}^{m}\ket{\phi_{i,j}} \mid j=1,2,\ldots,n\}$ be a set of OPSs in $\mathcal{H}$. The local rank of $\mathcal{S}$ on the $i$-th subsystem, denoted by $r_i$, is defined as the rank of the vector set $\{\ket{\phi_{i,j}}\mid j=1,2,\ldots,n\}$.
	\end{definition}

	
	
	
	
	For four tripartite OPSs, the distribution of the six pairwise orthogonality relations leads to three possible categories, namely $(4,1,1)$, $(3,2,1)$, and $(2,2,2)$. Their local distinguishability is summarized in the following lemmas.
	
	\begin{lemma}
		\label{lem:tripartite-411-321}
		\cite{Xu2025FourMultipartiteOPS}
		Four tripartite OPSs with the vector of the numbers of pairwise orthogonal relations $(4,1,1)$ or $(3,2,1)$ can be perfectly distinguished by LOCC.
	\end{lemma}
	
\begin{figure}[htbp]
		\centering
		\FOPincludegraphics[width=0.70\linewidth]{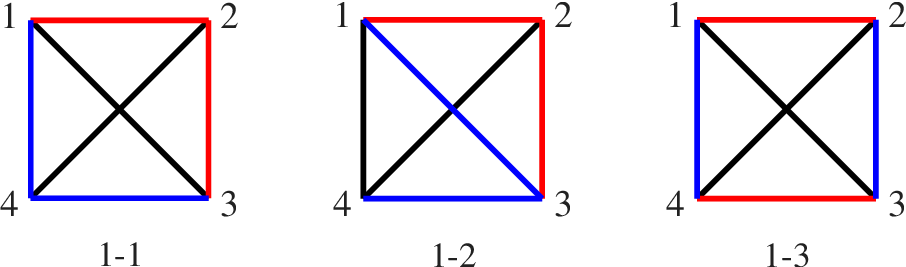}
		\caption{The feasible orthogonality graphs of four tripartite OPSs with the vector of the numbers of pairwise orthogonal relations $(2,2,2)$.}
		\label{fig:four-tripartite-222}
	\end{figure}

	For the category $(2,2,2)$, there are three different orthogonality graphs up to vertex relabeling and party permutation. We denote them by cases $(1\text{-}1)$, $(1\text{-}2)$, and $(1\text{-}3)$, as shown in Fig.~\ref{fig:four-tripartite-222}.
	
	\begin{lemma}
		\label{lem:tripartite-222-dist}\cite{Xu2025FourMultipartiteOPS}
		Four tripartite OPSs with the vector of the numbers of pairwise orthogonality relations $(2,2,2)$ can be perfectly distinguished by LOCC if their orthogonality graph corresponds to case $(1\text{-}1)$ or case $(1\text{-}2)$ up to vertex relabeling and party permutation, as shown in Fig.~\ref{fig:four-tripartite-222}.
	\end{lemma}
	
	\begin{lemma}
		\label{lem:tripartite-222-exception}\cite{Xu2025FourMultipartiteOPS}
		For four tripartite OPSs with the vector of the numbers of pairwise orthogonality relations $(2,2,2)$, if their orthogonality graph corresponds to case $(1\text{-}3)$ in 		Fig.~\ref{fig:four-tripartite-222} and the local ranks satisfy $r_1=r_2=r_3=2$, then they cannot be perfectly distinguished by LOCC. In the other local-rank cases, such a set can be distinguished by LOCC with a nonzero probability.
	\end{lemma}
	
	\begin{lemma}
		\label{lem:four-ops-multipartite}\cite{Xu2025FourMultipartiteOPS}	
		For $m\in\{4,5,6\}$, any set of four $m$-partite OPSs, any two of which are orthogonal on exactly one subsystem, can be perfectly distinguished by LOCC if each subsystem contains at least one pairwise orthogonality relation.	
	\end{lemma}

	The following four lemmas will be used in the subsequent analysis of five five-partite OPSs.

	\begin{lemma}
		\label{lem:degree-four}\cite{Xu2026SixBipartiteOPS}
		Let $\mathcal{S}$ be a set of five  OPSs, any two of which are orthogonal on exactly one subsystem, and the orthogonality graph of $S$ is $G=(V,\,E_1\cup E_2\cup E_3\cup E_4\cup E_5)$. Suppose that there exists a vertex $v\in V$ and a subsystem $p$ such that $\deg_p(v)=4$. Then the state corresponding to $v$ can be exactly identified by LOCC.
	\end{lemma}
	
	\begin{lemma}
		\label{lem:degree-three-one}\cite{Xu2026SixBipartiteOPS}
		Let $\mathcal{S}$ be a set of five  OPSs, any two of which are orthogonal on exactly one subsystem, and the orthogonality graph of $S$ is $G=(V,\,E_1\cup E_2\cup E_3\cup E_4\cup E_5)$. Suppose that there exists a vertex $v\in V$ and two distinct subsystems $p$ and $q$ such that $\deg_p(v)=3$ and $\deg_q(v)=1$. Then the state corresponding to $v$ can be distinguished by LOCC.
	\end{lemma}
	
	\begin{lemma}
		\label{lem:degree-two-two}\cite{Xu2026FiveQuadripartiteOPS}
		Let $\mathcal{S}$ be a set of five OPSs, any two of which are orthogonal on exactly one subsystem. Suppose that there exists a vertex $v$ in the orthogonality graph of $S$ and two distinct subsystems $p$ and $q$ such that $\deg_p(v)=\deg_q(v)=2$. Let \(N_p(v)=\{u_1,u_2\}\) and \(N_q(v)=\{w_1,w_2\}\). If the edge between $w_1$ and $w_2$ is not of color $p$, then the $p$-th party can perform an orthogonality-preserving local measurement on the $p$-th subsystem such that the state corresponding to $v$ can be identified by LOCC. 
	\end{lemma}
	
	\begin{lemma}
		\label{lem:degree-two-one-one}\cite{Xu2026FiveQuadripartiteOPS}
		Let $S$ be a set of five OPSs, any two of which are orthogonal on exactly one subsystem. Suppose that there exists a vertex $v$ in the orthogonality graph of $S$ and three distinct subsystems $p$, $q$, and $r$ such that $\deg_p(v)=2$ and $\deg_q(v)=\deg_r(v)=1$. Let $N_p(v)=\{u_1,u_2\}$, $N_q(v)=\{w_1\}$, and $N_r(v)=\{w_2\}$. If the edge between $w_1$ and $w_2$ is not of color $p$, then the $p$-th party can perform an orthogonality-preserving local measurement on the $p$-th subsystem such that the state corresponding to $v$ can be identified by LOCC.
	\end{lemma}
	
	\section{Local Distinguishability of Five Five-Partite OPSs}
	\label{sec:distinguishable}

	In this section, we investigate the local distinguishability of five five-partite OPSs, where any two of the five states are orthogonal on exactly one subsystem. Thus, there are ten pairwise orthogonality relations distributed among five subsystems.
	
	For the genuinely five-partite cases considered here, each subsystem contains at least one pairwise orthogonality relation. Arranging the five components of the vector defined in Sec.~\ref{sec:preliminaries} in nonincreasing order, up to permutation of the subsystems, we obtain seven categories, namely $(6,1,1,1,1)$, $(5,2,1,1,1)$, $(4,3,1,1,1)$, $(4,2,2,1,1)$, $(3,3,2,1,1)$, $(3,2,2,2,1)$, and $(2,2,2,2,2)$.
	
	For each of the seven categories, the pairwise orthogonality relations among five OPSs are represented by colored orthogonality graphs. Two orthogonality graphs are regarded as identical if one can be transformed into the other by relabeling the vertices and permuting the five subsystems.
	
	After removing graphs regarded as identical under these transformations, we obtain $443$ different orthogonality graphs. The numbers of graphs corresponding to $(6,1,1,1,1)$, $(5,2,1,1,1)$, $(4,3,1,1,1)$, $(4,2,2,1,1)$, $(3,3,2,1,1)$, $(3,2,2,2,1)$, and $(2,2,2,2,2)$ are $6$, $30$, $47$, $98$, $122$, $125$, and $15$, respectively. The sets of five OPSs corresponding to the graphs in the first three categories can be perfectly distinguished by LOCC, while the remaining four categories contain some exceptional cases.
	
\begin{figure*}[htbp]
		\centering
		\FOPincludegraphics[width=0.80\linewidth]{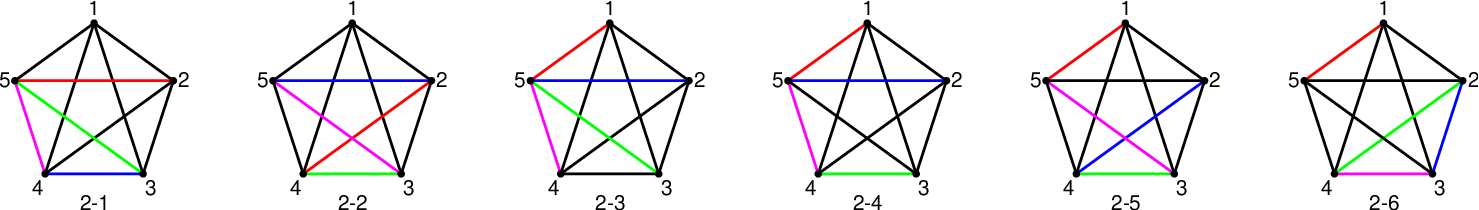}
		\caption{\centering All feasible orthogonality graphs for category $(6,1,1,1,1)$.}
		\label{fig:61111}
	\end{figure*}

\begin{figure*}[htbp]
		\centering
		\FOPincludegraphics[width=0.80\linewidth]{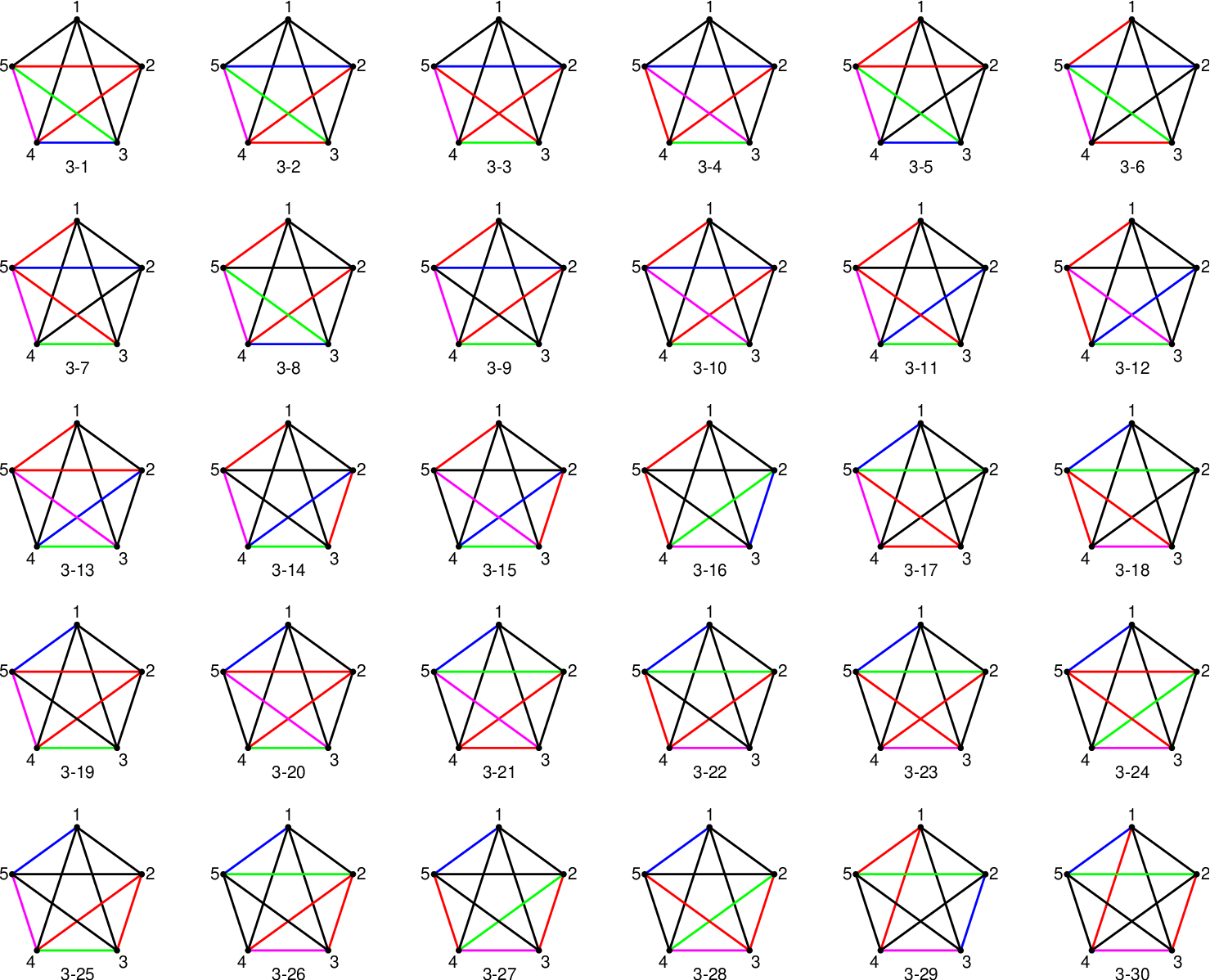}
		\caption{\centering All feasible orthogonality graphs for category $(5,2,1,1,1)$.}
		\label{fig:52111}
	\end{figure*}

\begin{figure*}[htbp]
		\centering
		
		\FOPincludegraphics[width=0.80\linewidth]{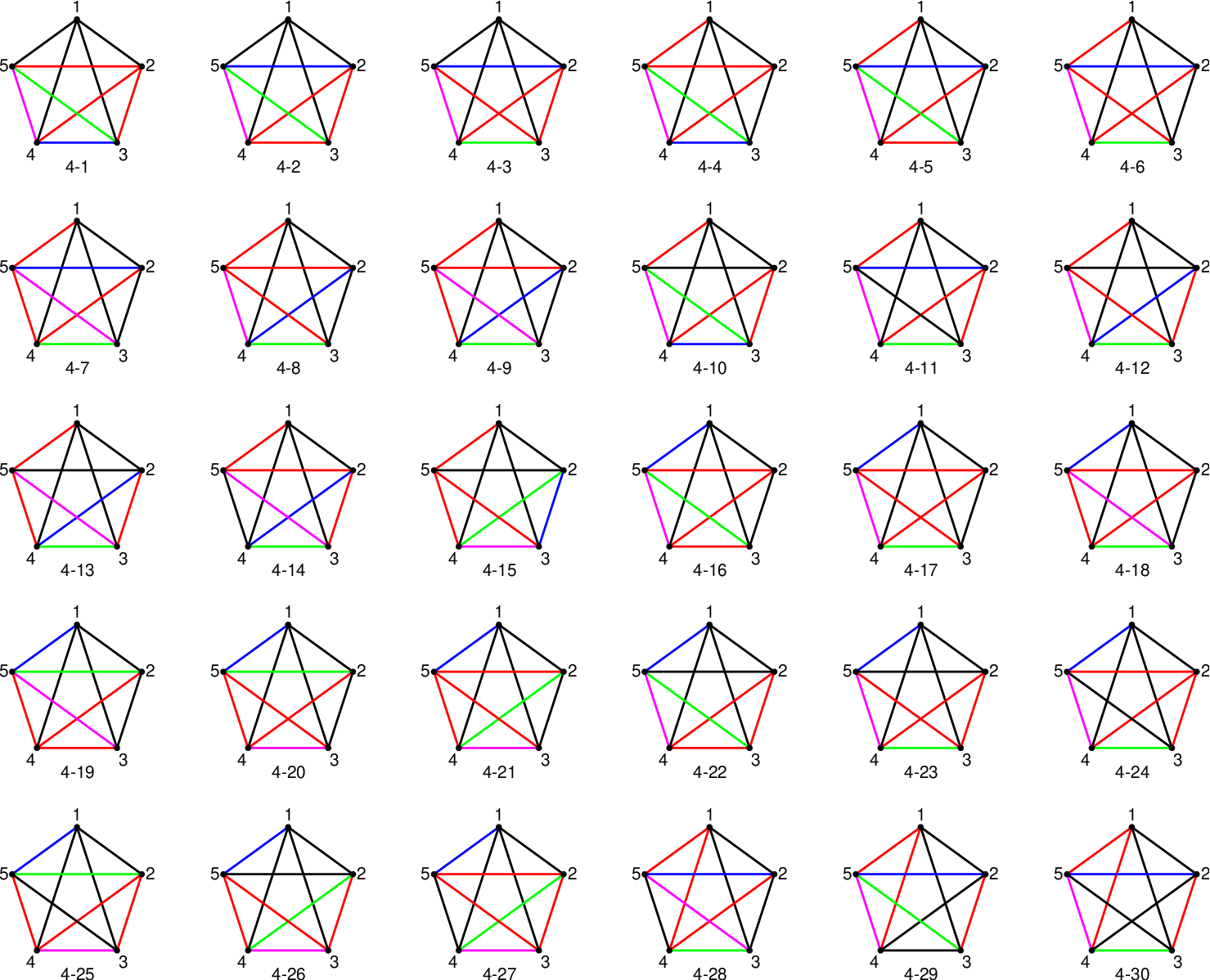}
		
		\vspace{2mm}
		
		\FOPincludegraphics[width=0.80\linewidth]{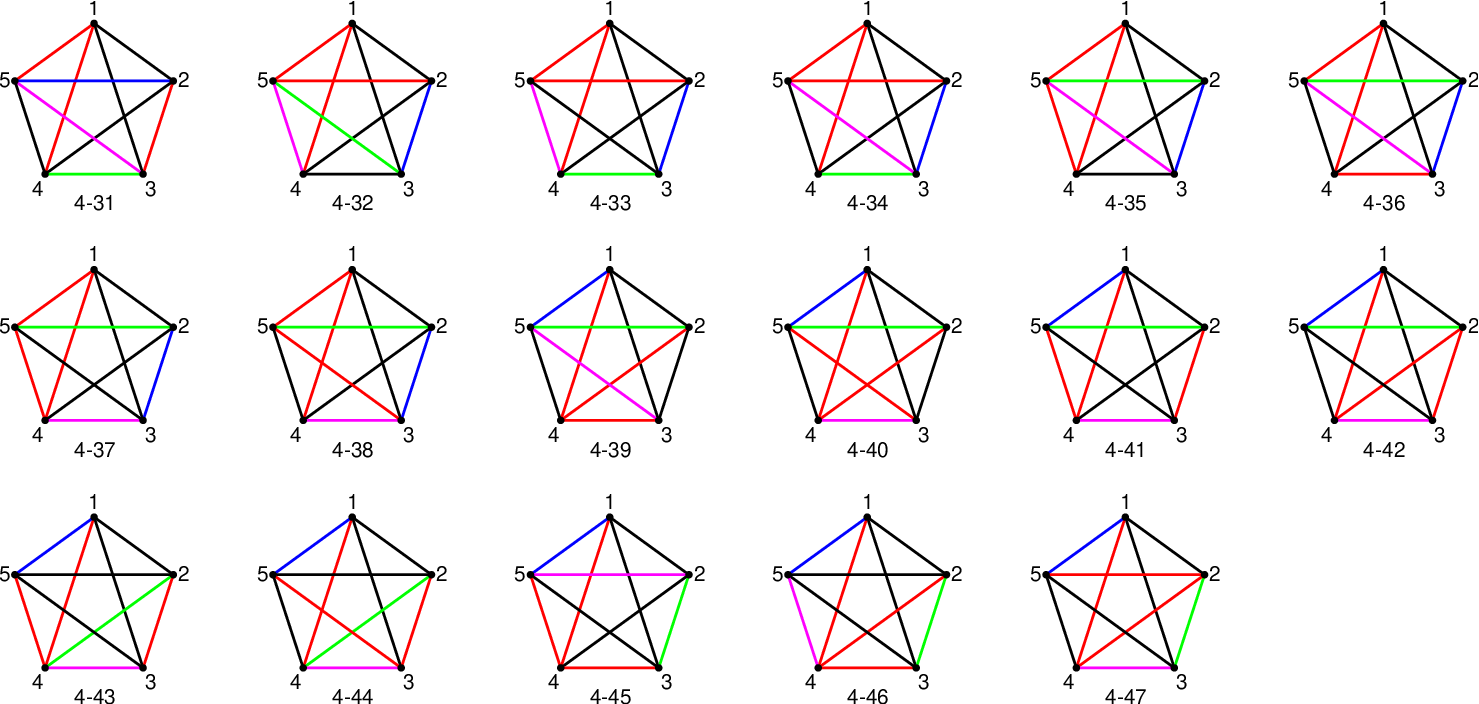}
		
		\caption{\centering All feasible orthogonality graphs for category $(4,3,1,1,1)$.}
		\label{fig:43111}
	\end{figure*}

	We first consider the three categories, namely $(6,1,1,1,1)$, $(5,2,1,1,1)$, and $(4,3,1,1,1)$. All different orthogonality graphs for these three categories are shown in Figs.~\ref{fig:61111}, \ref{fig:52111}, and \ref{fig:43111}.
	
\begin{figure}[htbp]
		\centering
		\FOPincludegraphics[width=0.80\linewidth]{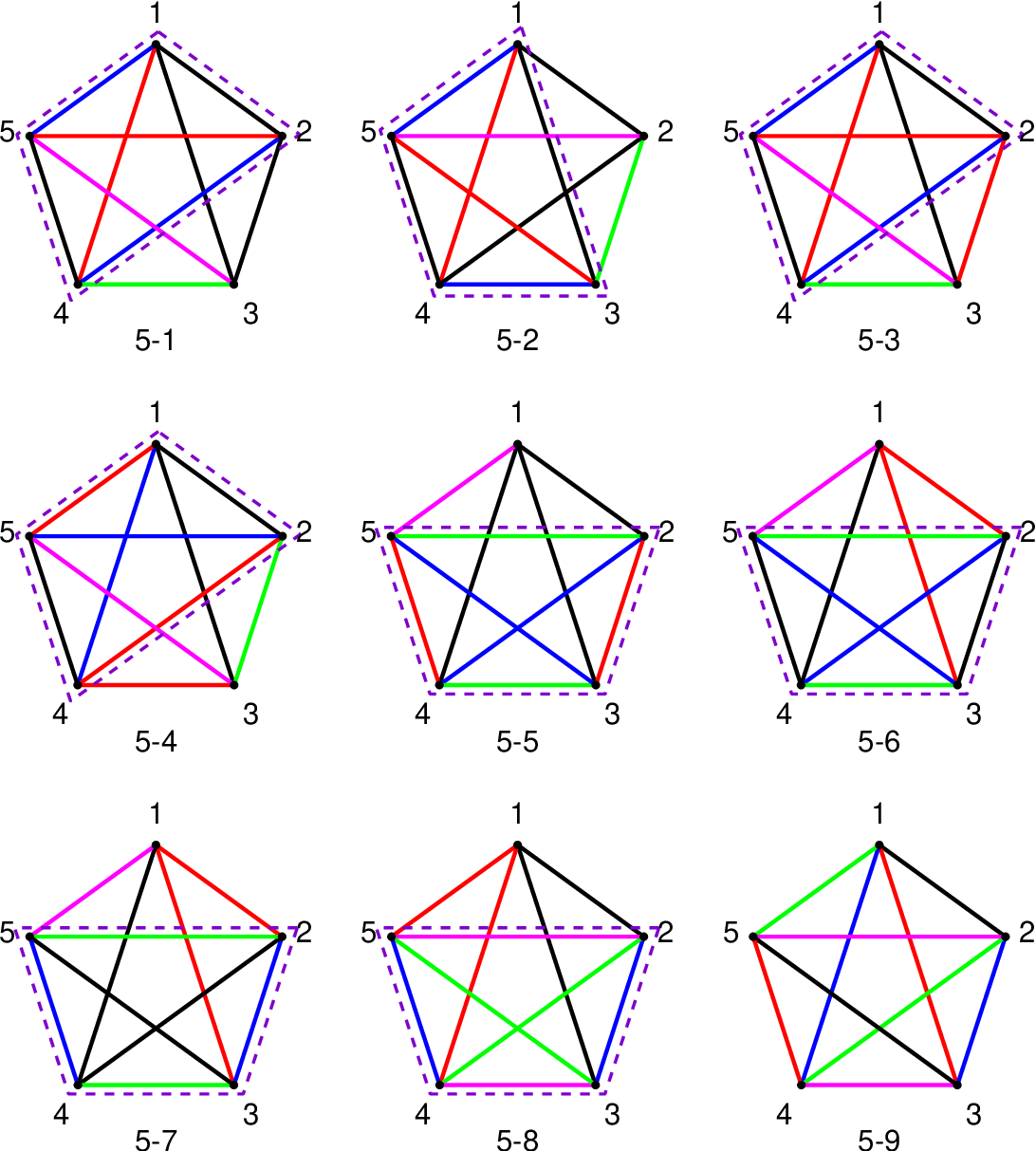}
		\caption{\centering The nine exceptional orthogonality graphs in Theorem~\ref{thm:main-distinguishability}.}
		\label{fig:nine-exceptional}
	\end{figure}

	The remaining four categories contain $98$, $122$, $125$, and $15$ different orthogonality graphs, respectively. Nine of these graphs are exceptional, as shown in Fig.~\ref{fig:nine-exceptional}. Cases $(5\text{-}1)$ and $(5\text{-}2)$ belong to category $(4,2,2,1,1)$; cases $(5\text{-}3)$ and $(5\text{-}4)$ belong to category $(3,3,2,1,1)$; cases $(5\text{-}5)$--$(5\text{-}7)$ belong to category $(3,2,2,2,1)$; and cases $(5\text{-}8)$ and $(5\text{-}9)$ belong to category $(2,2,2,2,2)$.
	
\begin{figure*}[htbp]
		\centering
		
		\FOPincludegraphics[width=0.80\linewidth]{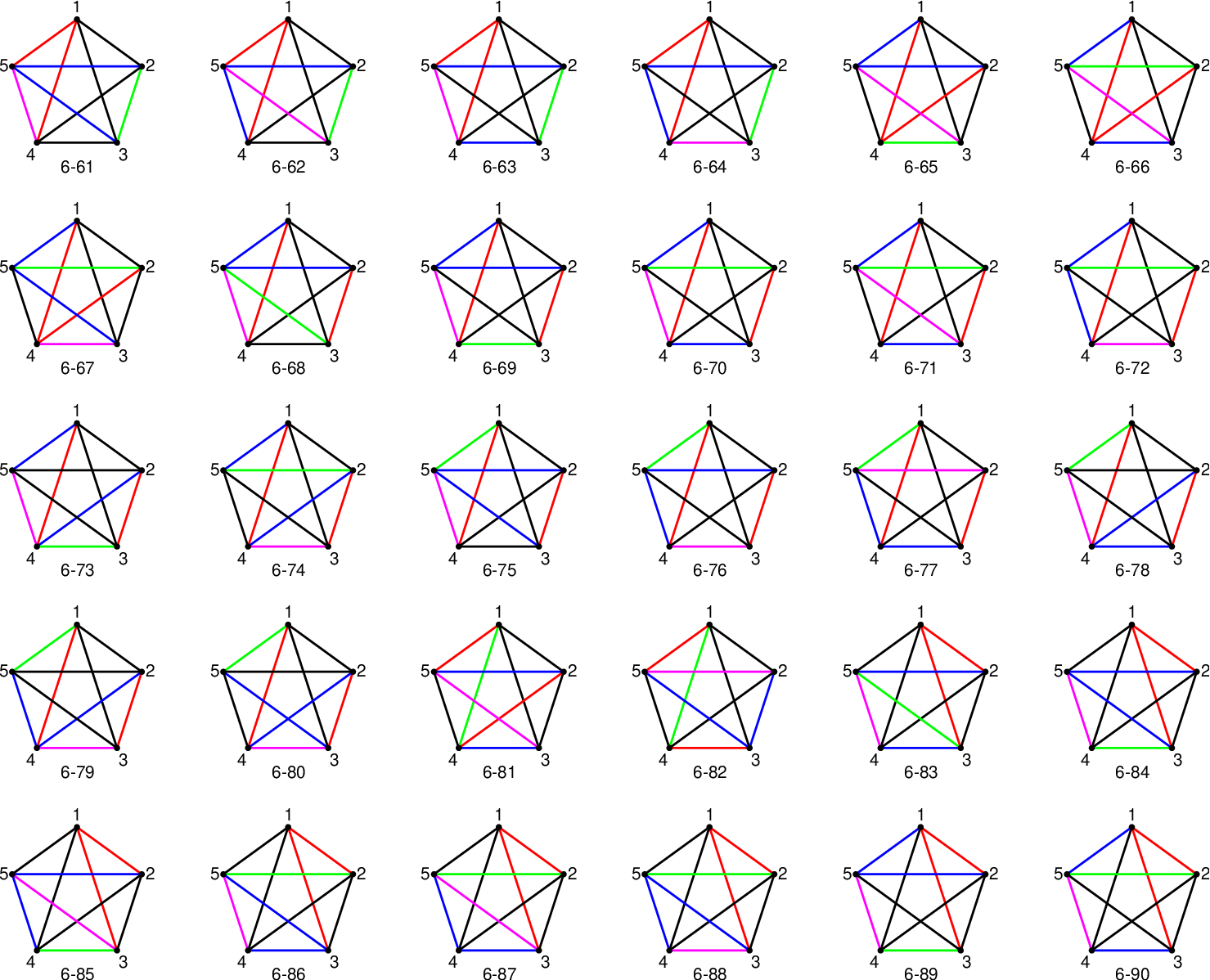}
		
		\vspace{1mm}
		
		\FOPincludegraphics[width=0.80\linewidth]{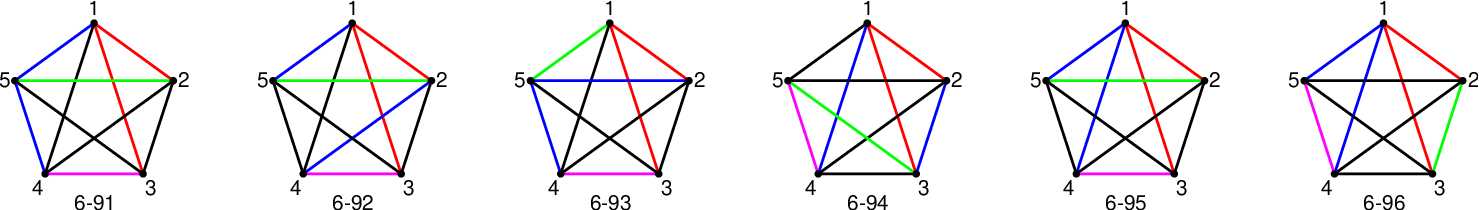}
		
		\caption{\centering Orthogonality graphs for the 96 cases perfectly distinguishable by LOCC in category $(4,2,2,1,1)$.}
		\label{fig:42211-2}
	\end{figure*}

\begin{figure*}[htbp]
		\centering
		
		\FOPincludegraphics[width=0.80\linewidth]{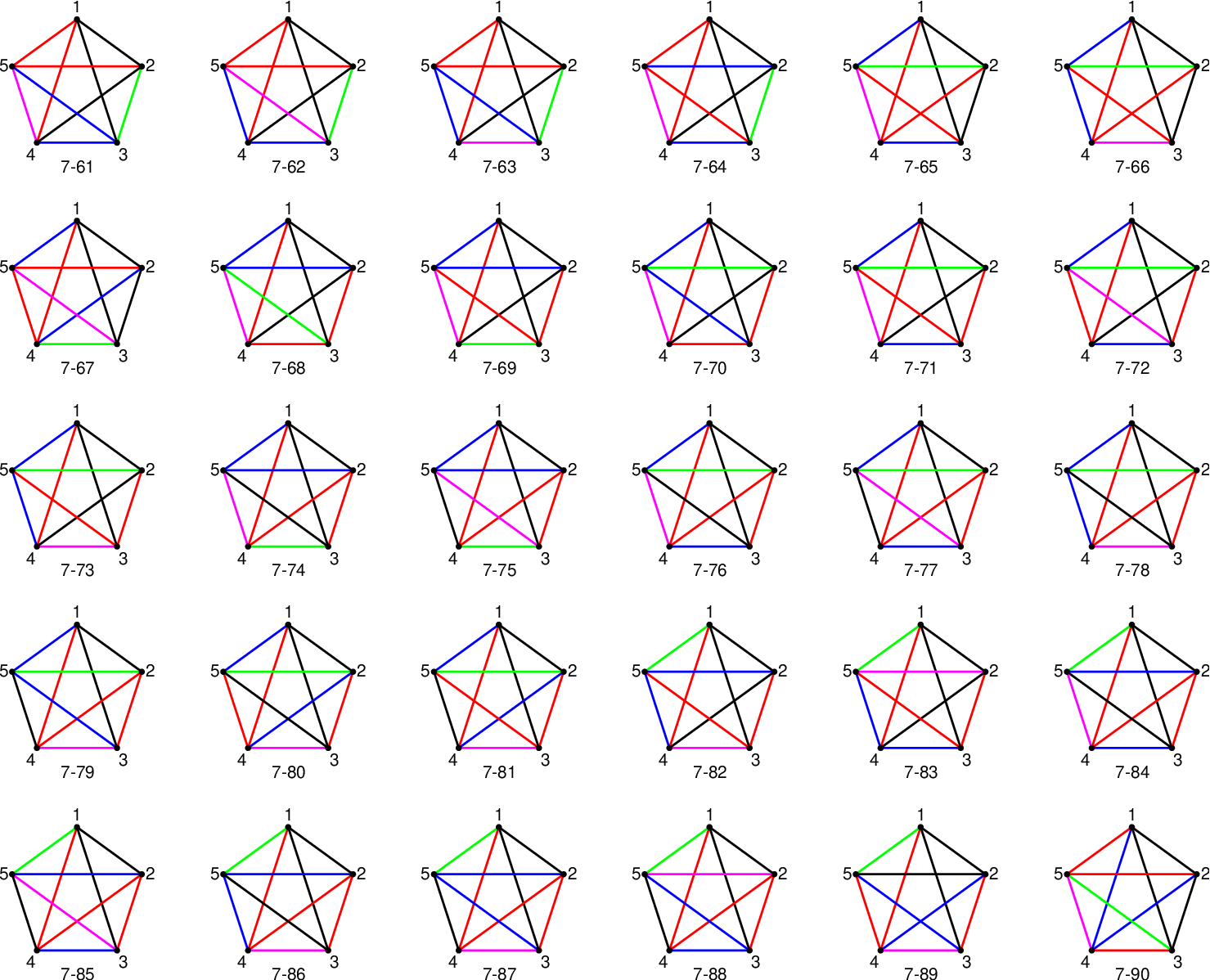}
		
		\vspace{1mm}
		
		\FOPincludegraphics[width=0.80\linewidth]{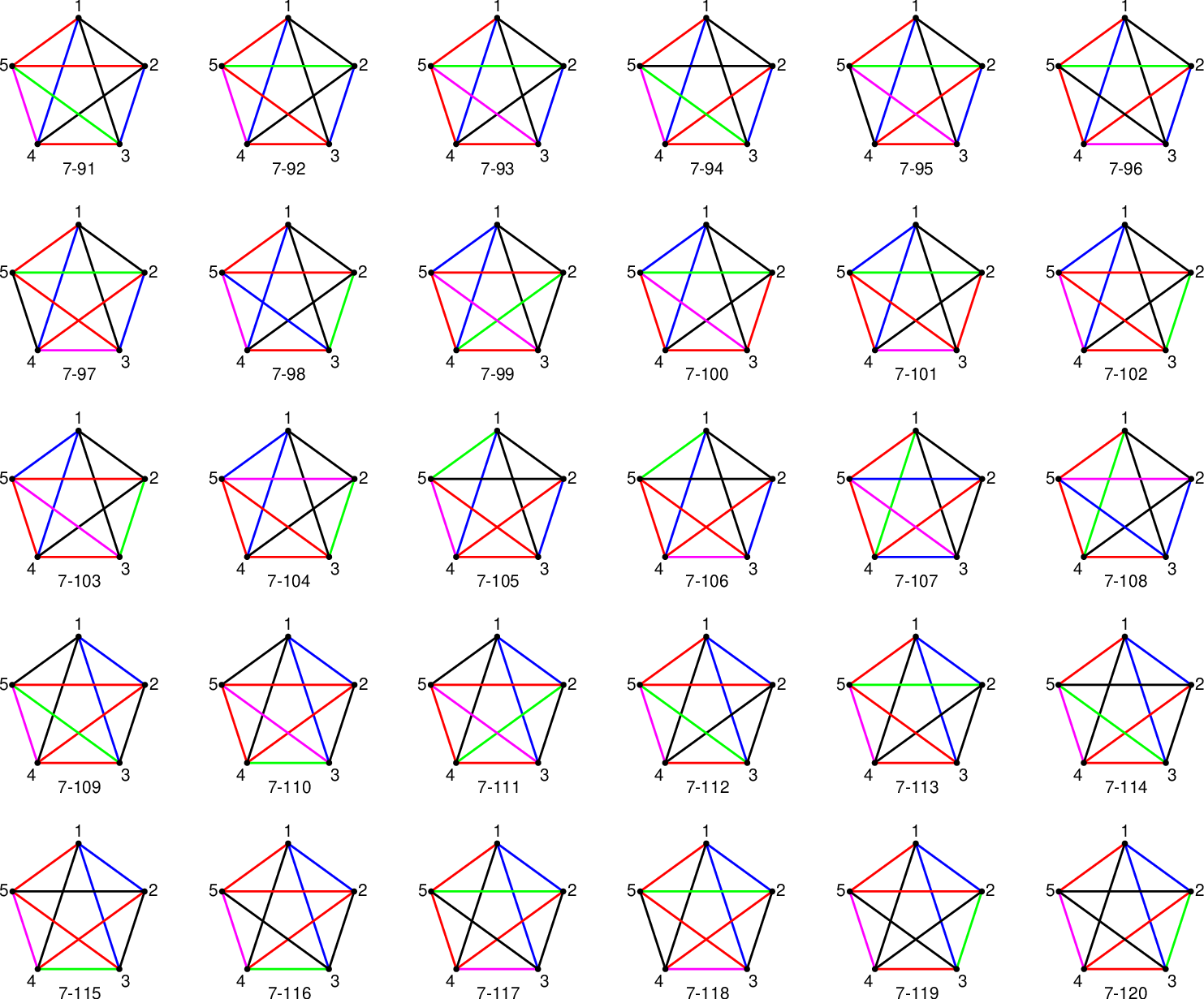}
		
		\caption{\centering Orthogonality graphs for the 120 cases perfectly distinguishable by LOCC in category $(3,3,2,1,1)$.}
		\label{fig:33211-2}
	\end{figure*}

\begin{figure*}[htbp]
		\centering
		
		\FOPincludegraphics[width=0.75\linewidth]{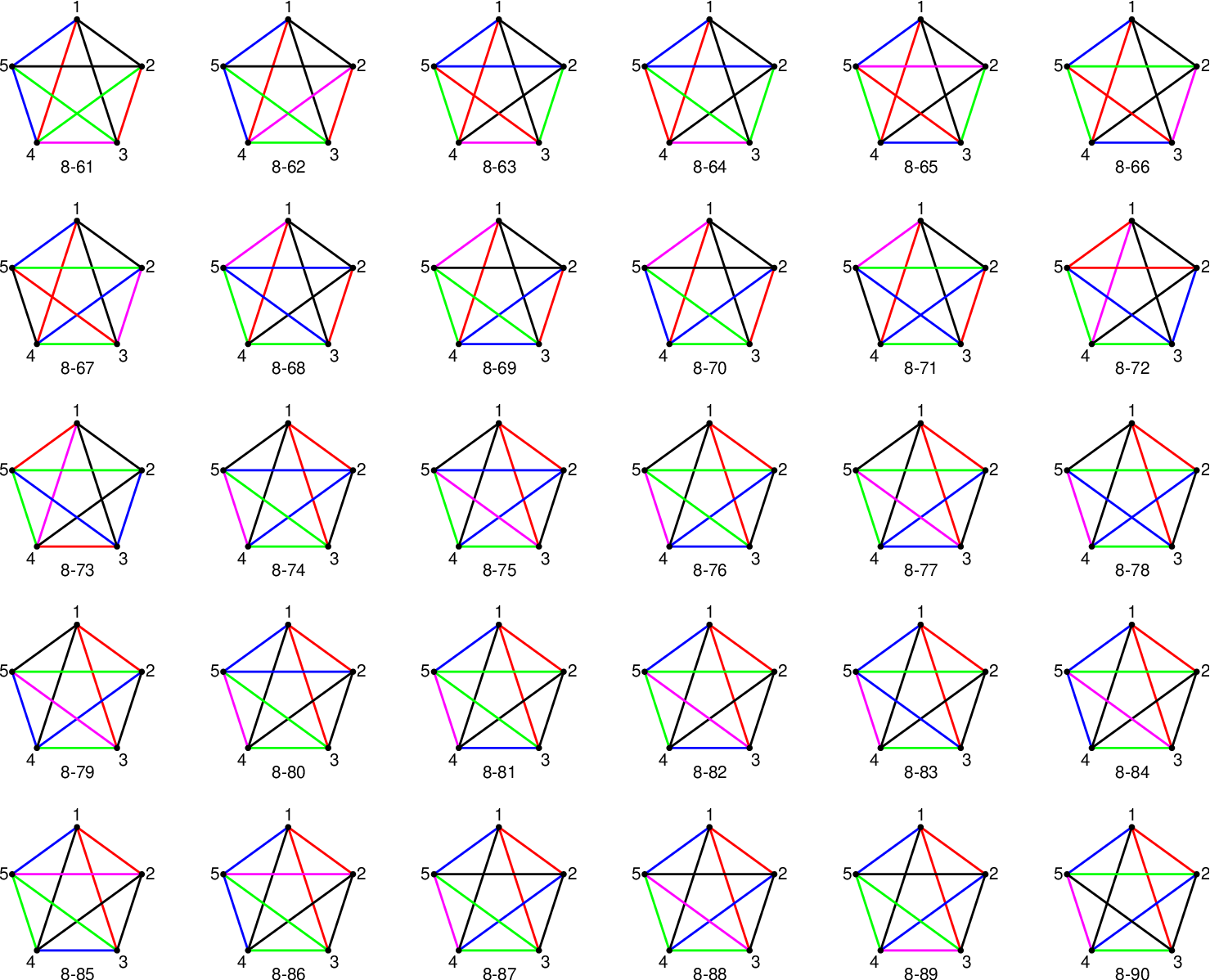}
		
		\vspace{1mm}
		
		\FOPincludegraphics[width=0.75\linewidth]{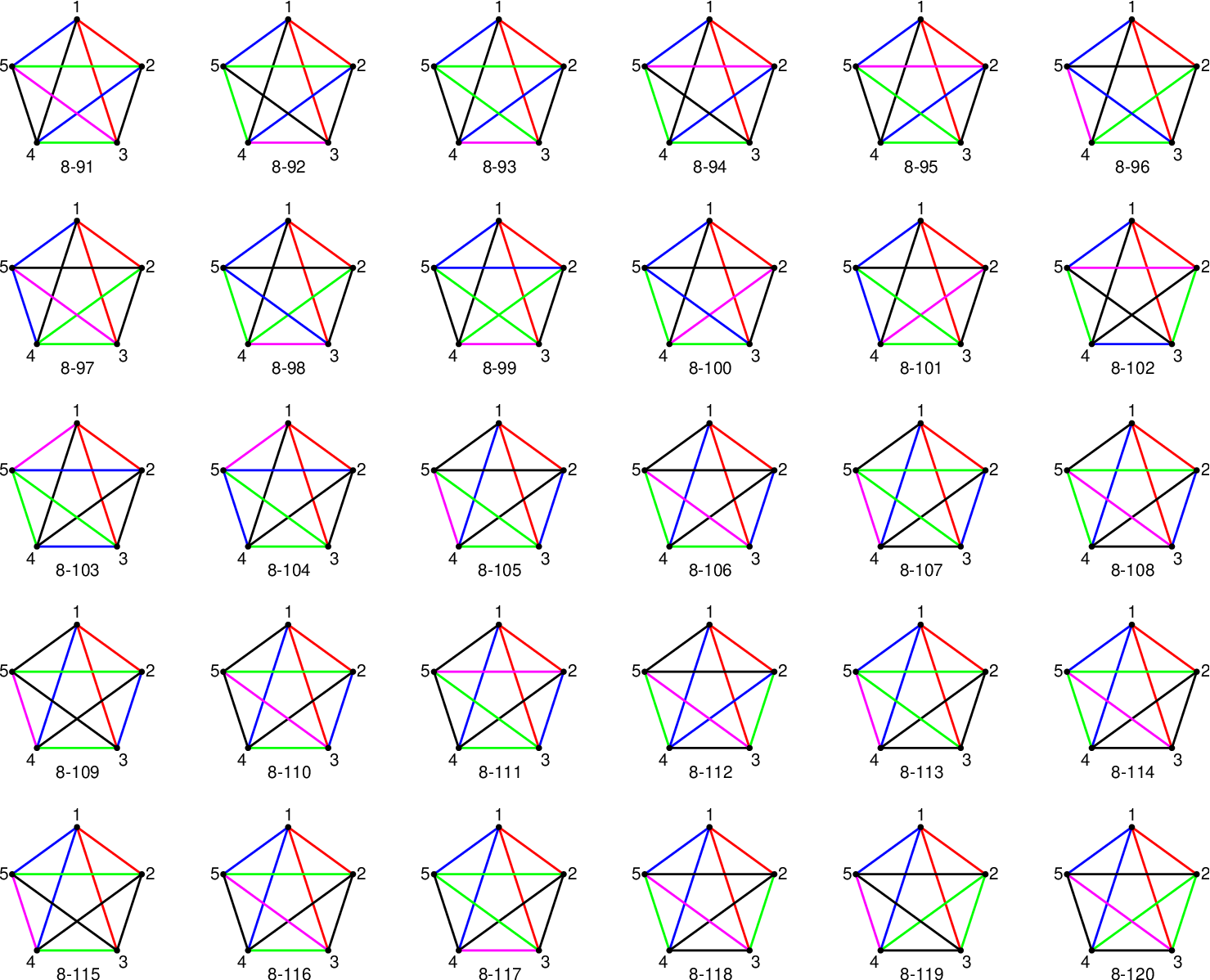}
		
		\vspace{1mm}
		
		\makebox[\linewidth][l]{\hspace*{2.3cm}\FOPincludegraphics[width=0.23\linewidth]{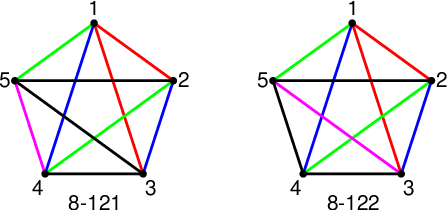}}
		
		\caption{\centering Orthogonality graphs for the 122 cases perfectly distinguishable by LOCC in category $(3,2,2,2,1)$.}
		\label{fig:32221-2}
	\end{figure*}

\begin{figure*}[htbp]
		\centering
		
		\FOPincludegraphics[width=0.80\linewidth]{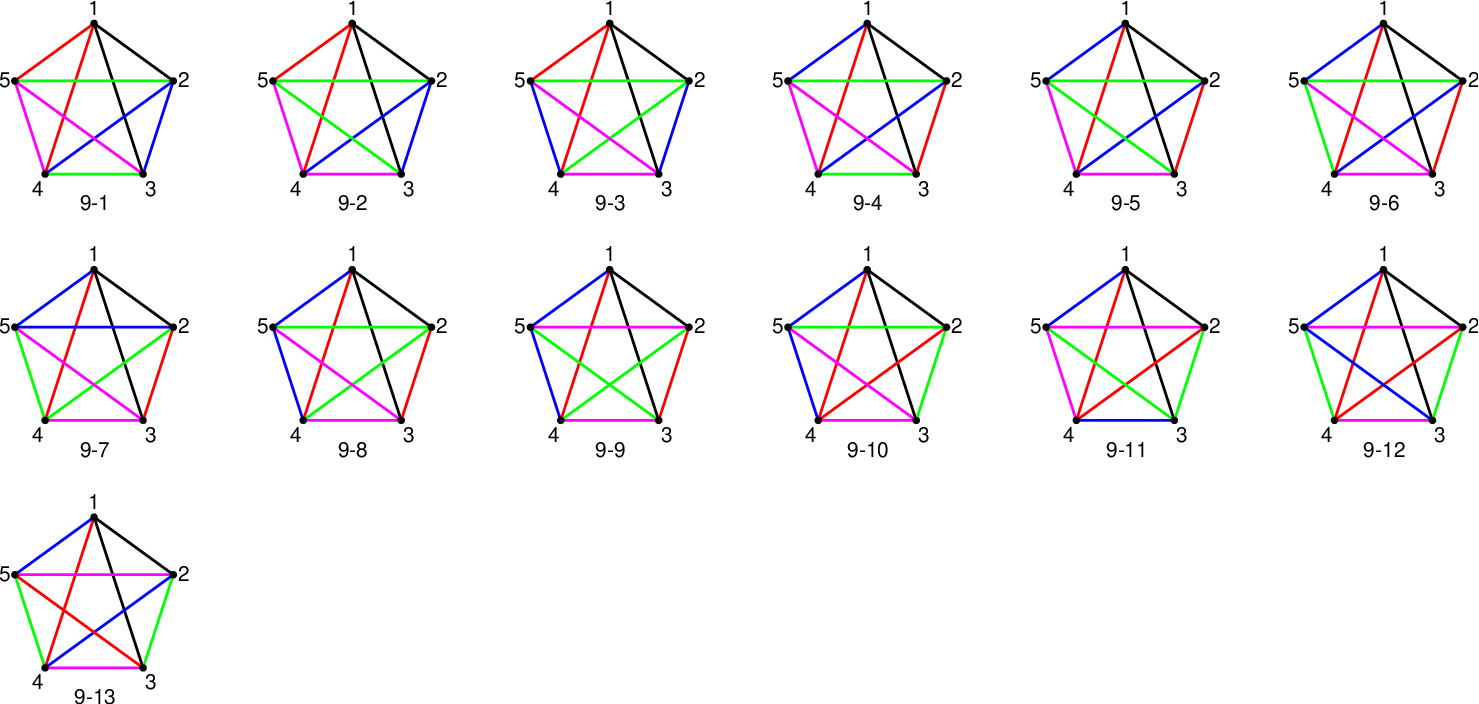}
		
		\caption{\centering Orthogonality graphs for the 13 cases perfectly distinguishable by LOCC in category $(2,2,2,2,2)$.}
		\label{fig:22222}
	\end{figure*}

	The sets of five OPSs corresponding to the remaining $96$, $120$, $122$, and $13$ graphs in these four categories can be perfectly distinguished by LOCC. The corresponding orthogonality graphs are shown in Figs.~\ref{fig:42211-2}, \ref{fig:33211-2}, \ref{fig:32221-2}, and \ref{fig:22222}.
	
	\begin{figure*}[htbp]
		\centering
		
		\FOPincludegraphics[width=0.80\linewidth]{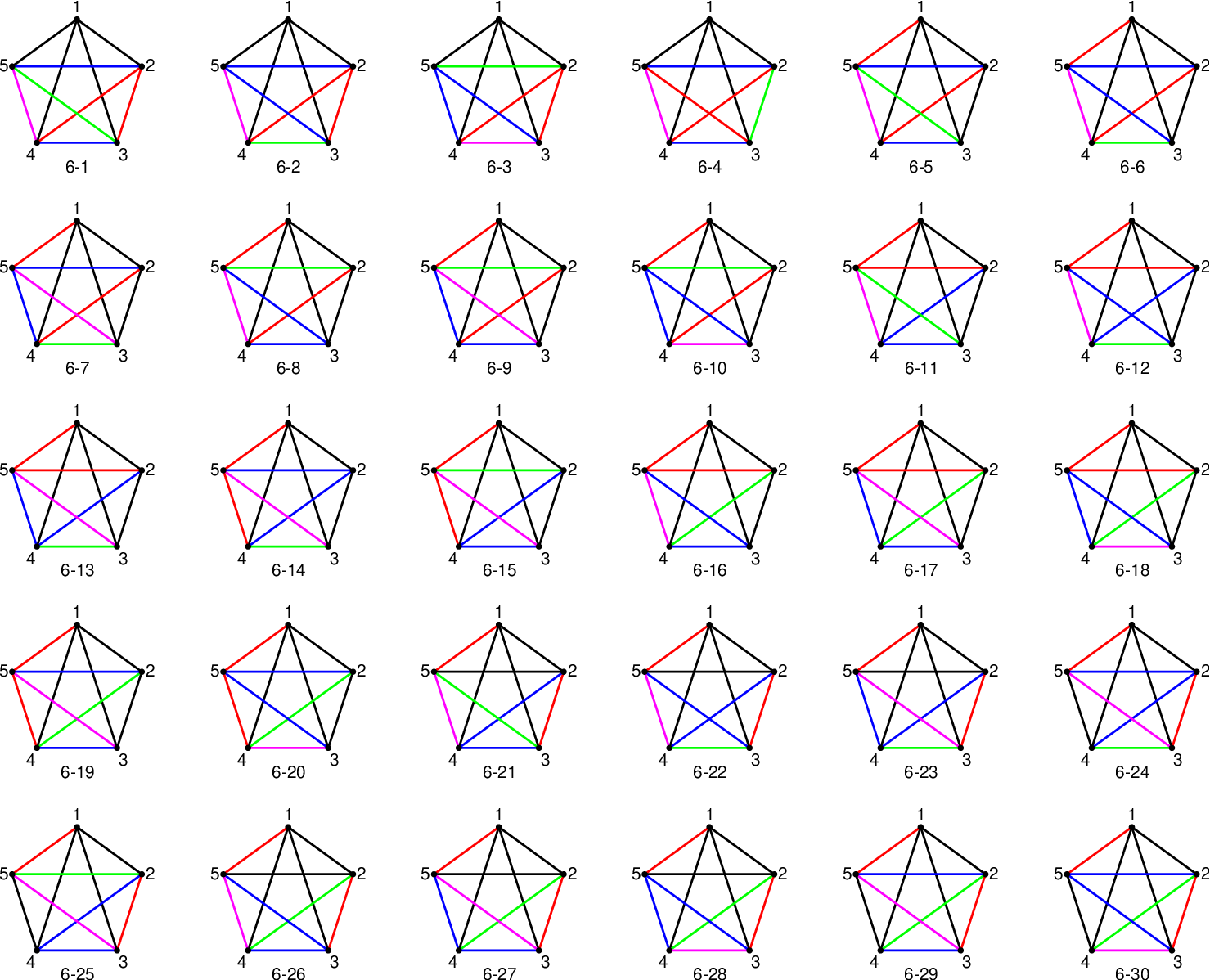}
		
		\vspace{1mm}
		
		\FOPincludegraphics[width=0.80\linewidth]{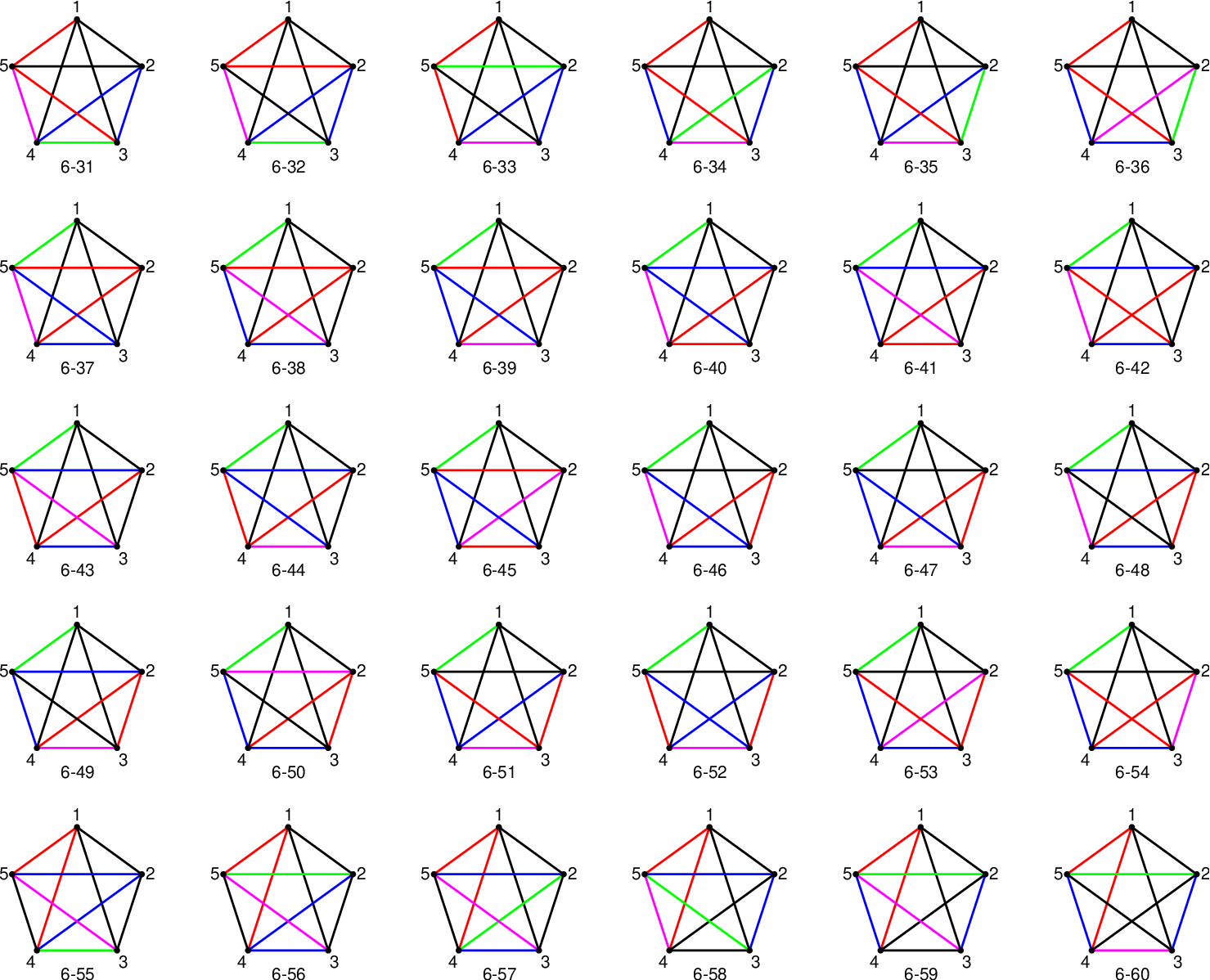}
		
		\label{fig:42211-1}
	\end{figure*}
	
	\begin{figure*}[htbp]
		\centering
		
		\FOPincludegraphics[width=0.80\linewidth]{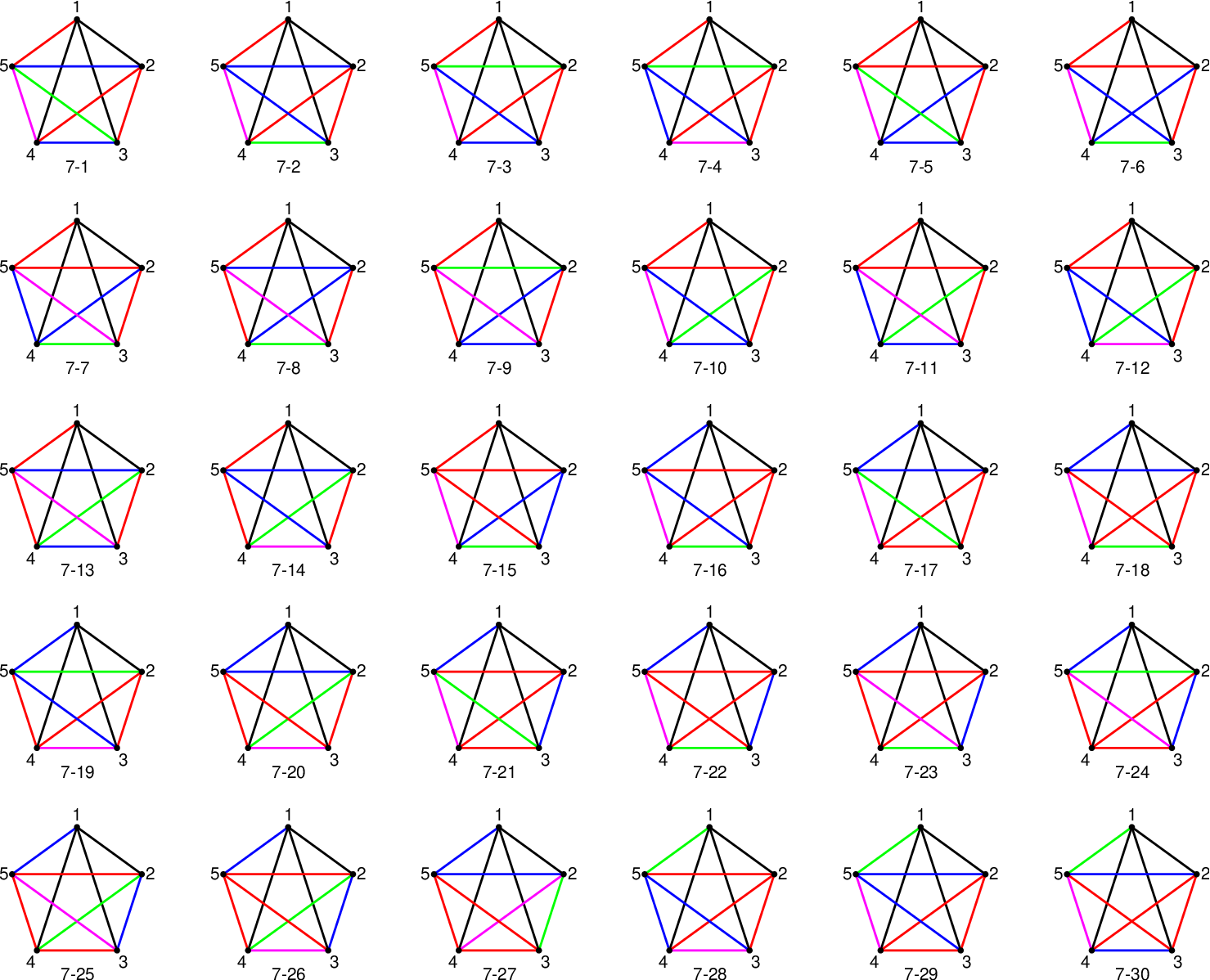}
		
		\vspace{1mm}
		
		\FOPincludegraphics[width=0.80\linewidth]{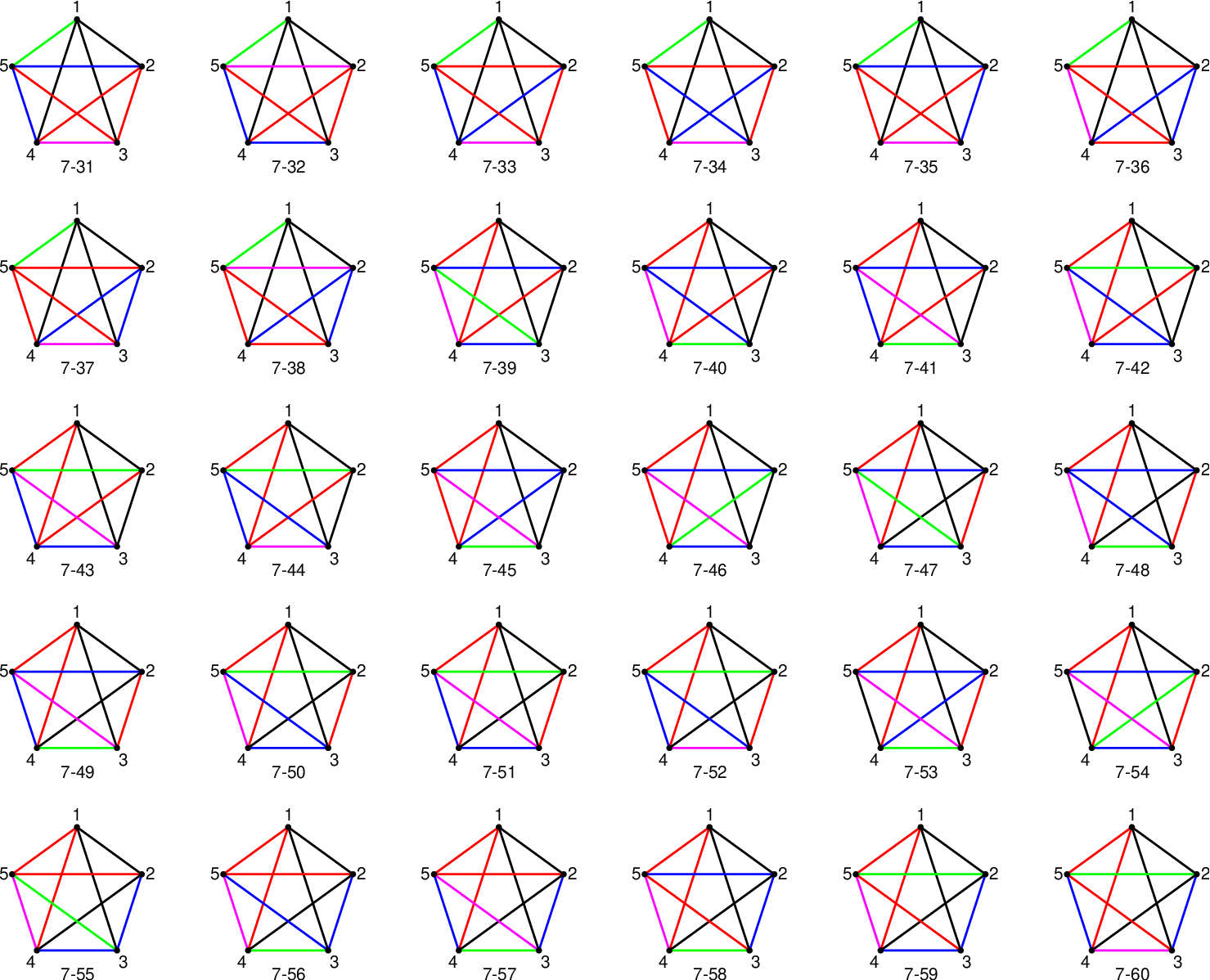}
		
		\label{fig:33211-1}
	\end{figure*}
	
	\begin{figure*}[htbp]
		\centering
		
		\FOPincludegraphics[width=0.80\linewidth]{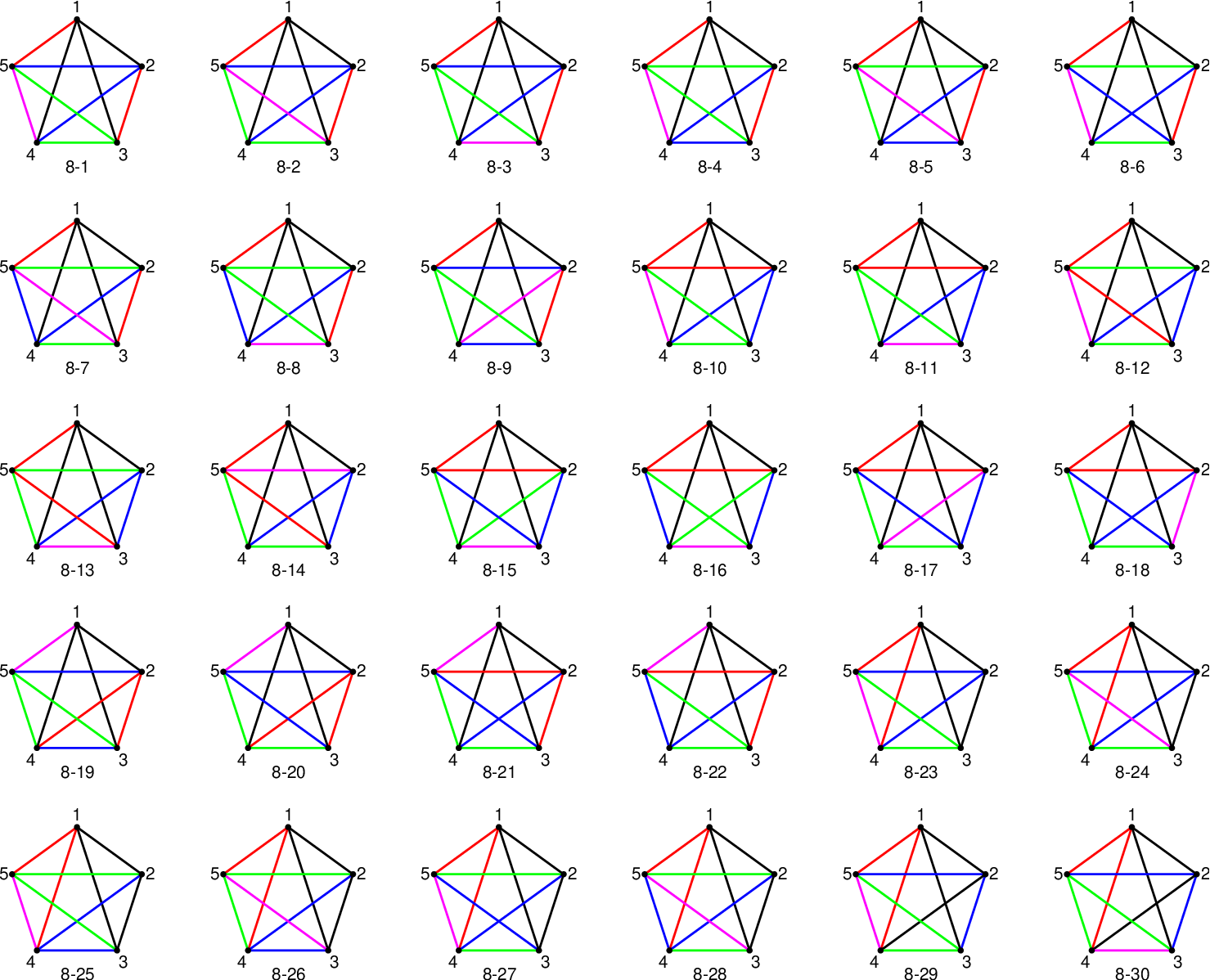}
		
		\vspace{1mm}
		
		\FOPincludegraphics[width=0.80\linewidth]{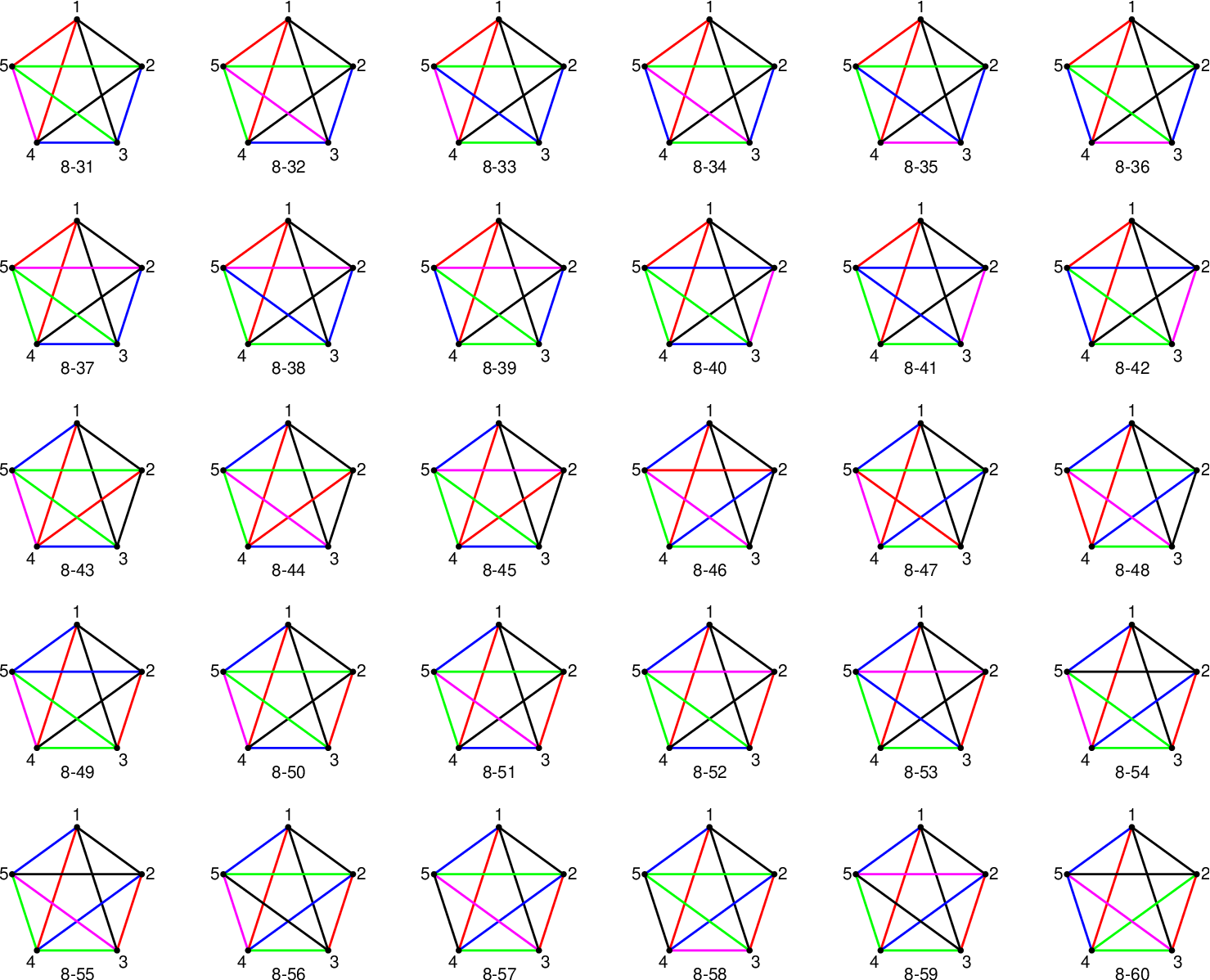}
		
		\label{fig:32221-1}
	\end{figure*}
	
	\begin{table*}[htbp]
		\centering
		\caption{All feasible orthogonality configurations of five five-partite OPSs.}
		\label{tab:distinguishability-summary}
		
		\renewcommand{\arraystretch}{1.15}
		\setlength{\tabcolsep}{8pt}
		
		\begin{tabular}{cccc}
			\toprule
			Category
			&
			\shortstack{Distinguishable\\cases}
			&
			\shortstack{Exceptional\\cases}
			&
			\shortstack{Total\\cases}
			\\
			\midrule
			
			$(6,1,1,1,1)$ & 6   & 0 & 6   \\
			
			$(5,2,1,1,1)$ & 30  & 0 & 30  \\
			
			$(4,3,1,1,1)$ & 47  & 0 & 47  \\
			
			$(4,2,2,1,1)$ & 96  & 2 & 98  \\
			
			$(3,3,2,1,1)$ & 120 & 2 & 122 \\
			
			$(3,2,2,2,1)$ & 122 & 3 & 125 \\
			
			$(2,2,2,2,2)$ & 13  & 2 & 15  \\
			
			\midrule
			Total & 434 & 9 & 443 \\
			\bottomrule
		\end{tabular}
	\end{table*}
	
	Table~\ref{tab:distinguishability-summary} summarizes the local distinguishability of five five-partite OPSs corresponding to the $443$ different orthogonality graphs in the seven genuinely five-partite categories. The sets of five OPSs corresponding to the graphs in the first three categories can be perfectly distinguished by LOCC. For the remaining four categories, there exist nine exceptional cases that require separate analysis. Thus, the sets of five OPSs corresponding to $434$ of the $443$ graphs can be perfectly distinguished by LOCC.

	The perfect distinguishability of the sets of five OPSs corresponding to the 434 nonexceptional graphs can be established using orthogonality-preserving local measurements. By applying Lemmas~\ref{lem:degree-four}, \ref{lem:degree-three-one}, \ref{lem:degree-two-two}, and \ref{lem:degree-two-one-one}, the five-state discrimination problem can be reduced to problems involving fewer states. The local distinguishability of the remaining states can then be determined using the results introduced in Sec.~\ref{sec:preliminaries}.
	
	The cases that can be treated using similar arguments are discussed together. The detailed proof for the 434 cases is given in Appendix~\ref{app:proof-main}. The remaining nine exceptional cases are considered separately in Sec.~\ref{sec:exceptional}.
	
	
	\begin{theorem}
		\label{thm:main-distinguishability}
		Let $\mathcal{S}$ be a set of five five-partite OPSs, where any two of these states are orthogonal on exactly one subsystem. $\mathcal{S}$ can be perfectly distinguished by LOCC except for the following two types:
		
		\textnormal{(1)} There exist four states among the five OPSs and three distinct subsystems $p$, $q$, and $r$ such that, in the orthogonality graph of these four states, $\deg_p(v_j)=\deg_q(v_j)=\deg_r(v_j)=1$ for $j=1,2,3,4$.
		
		\textnormal{(2)} The five vertices can be labeled $v_1,\ldots,v_5$ such that $\deg_j(v_j)=0$ and $\deg_p(v_j)=1$ for every $p\neq j$, where $j,p\in\{1,2,3,4,5\}$.
	\end{theorem}
	
	The first type consists of cases $(5\text{-}1)$--$(5\text{-}8)$, while the second type corresponds to case $(5\text{-}9)$ in Fig.~\ref{fig:nine-exceptional}.

	
	\section{Local Indistinguishability of the Exceptional Cases}
	\label{sec:exceptional}

	In this section, we discuss the local distinguishability of five five-partite OPSs corresponding to the nine exceptional orthogonality graphs in Theorem~\ref{thm:main-distinguishability}. These exceptional cases are divided into two types and analyzed separately.
	
	We first consider the eight exceptional cases of the first type, labeled $(5\text{-}1)$--$(5\text{-}8)$ in Fig.~\ref{fig:nine-exceptional}.
	
	\textnormal{(1)} For case $(5\text{-}1)$, the only orthogonality relation on the fourth subsystem is between states $3$ and $4$. Let $\ket{\delta_3}$ denote the normalized fourth local component of state $3$. The fourth party performs the measurement $\{M,I-M\}$, where $M=\ket{\delta_3}\bra{\delta_3}$.
	
	For the outcome corresponding to $M$, state $4$ is excluded, leaving states $\{1,2,3,5\}$. Their vector of the numbers of pairwise orthogonal relations is $(3,1,1,0,1)$. Since there is no orthogonality relation on the fourth subsystem among these four states, their orthogonality relations correspond to the quadripartite vector $(3,1,1,1)$ after omitting the fourth subsystem. By Lemma~\ref{lem:four-ops-multipartite}, these four states can be perfectly distinguished by LOCC.
	
	For the outcome corresponding to $I-M$, state $3$ is excluded, leaving states $\{1,2,4,5\}$. Their vector of the numbers of pairwise orthogonal relations is $(2,2,2,0,0)$. Since there are no orthogonality relations on the fourth and fifth subsystems among these four states, the orthogonality graph induced by these four states is identical to case $(1\text{-}3)$ in Fig.~\ref{fig:four-tripartite-222} after an appropriate relabeling of the vertices and a permutation of subsystems. By Lemma~\ref{lem:tripartite-222-exception}, the corresponding tripartite OPSs cannot be perfectly distinguished by LOCC when their three local ranks are all equal to $2$, and can be distinguished by LOCC with nonzero probability otherwise.
	
	\textnormal{(2)} For case $(5\text{-}2)$, the only orthogonality relation on the fourth subsystem is between states $2$ and $3$. Let $\ket{\delta_2}$ denote the normalized fourth local component of state $2$. The fourth party performs the measurement $\{M,I-M\}$, where $M=\ket{\delta_2}\bra{\delta_2}$.
	
	For the outcome corresponding to $M$, state $3$ is excluded, leaving states $\{1,2,4,5\}$. Their vector of the numbers of pairwise orthogonal relations is $(3,1,1,0,1)$. Since there is no orthogonality relation on the fourth subsystem among these four states, their orthogonality relations correspond to the quadripartite vector $(3,1,1,1)$ after omitting the fourth subsystem. By Lemma~\ref{lem:four-ops-multipartite}, these four states can be perfectly distinguished by LOCC.

	For the outcome corresponding to $I-M$, state $2$ is excluded, leaving states $\{1,3,4,5\}$. Their vector of the numbers of pairwise orthogonal relations is $(2,2,2,0,0)$. Since there are no orthogonality relations on the fourth and fifth subsystems among these four states, the orthogonality graph induced by these four states, after omitting these two subsystems, is identical to case $(1\text{-}3)$ in Fig.~\ref{fig:four-tripartite-222} after an appropriate relabeling of the vertices and a permutation of subsystems. By Lemma~\ref{lem:tripartite-222-exception}, the corresponding tripartite OPSs cannot be perfectly distinguished by LOCC when their three local ranks are all equal to $2$, and can be distinguished by LOCC with nonzero probability otherwise.
	
	\textnormal{(3)} For case $(5\text{-}3)$, the only orthogonality relation on the fourth subsystem is between states $3$ and $4$. Let $\ket{\delta_3}$ denote the normalized fourth local component of state $3$. The fourth party performs the measurement $\{M,I-M\}$, where $M=\ket{\delta_3}\bra{\delta_3}$.
	
	For the outcome corresponding to $M$, state $4$ is excluded, leaving states $\{1,2,3,5\}$. Their vector of the numbers of pairwise orthogonal relations is $(2,2,1,0,1)$. Since there is no orthogonality relation on the fourth subsystem among these four states, their orthogonality relations correspond to the quadripartite vector $(2,2,1,1)$ after omitting the fourth subsystem. By Lemma~\ref{lem:four-ops-multipartite}, these four states can be perfectly distinguished by LOCC.
	
	For the outcome corresponding to $I-M$, state $3$ is excluded, leaving states $\{1,2,4,5\}$. Their vector of the numbers of pairwise orthogonal relations is $(2,2,2,0,0)$. Since there are no orthogonality relations on the fourth and fifth subsystems among these four states, the orthogonality graph induced by these four states, after omitting these two subsystems, is identical to case $(1\text{-}3)$ in Fig.~\ref{fig:four-tripartite-222} after an appropriate relabeling of the vertices and a permutation of subsystems. By Lemma~\ref{lem:tripartite-222-exception}, the corresponding tripartite OPSs cannot be perfectly distinguished by LOCC when their three local ranks are all equal to $2$, and can be distinguished by LOCC with nonzero probability otherwise.
	
	\textnormal{(4)} For case $(5\text{-}4)$, the only orthogonality relation on the fourth subsystem is between states $2$ and $3$. Let $\ket{\delta_3}$ denote the normalized fourth local component of state $3$. The fourth party performs the measurement $\{M,I-M\}$, where $M=\ket{\delta_3}\bra{\delta_3}$.
	
	For the outcome corresponding to $M$, state $2$ is excluded, leaving states $\{1,3,4,5\}$. Their vector of the numbers of pairwise orthogonal relations is $(2,2,1,0,1)$. Since there is no orthogonality relation on the fourth subsystem among these four states, their orthogonality relations correspond to the quadripartite vector $(2,2,1,1)$ after omitting the fourth subsystem. By Lemma~\ref{lem:four-ops-multipartite}, these four states can be perfectly distinguished by LOCC.
	
	For the outcome corresponding to $I-M$, state $3$ is excluded, leaving states $\{1,2,4,5\}$. Their vector of the numbers of pairwise orthogonal relations is $(2,2,2,0,0)$. Since there are no orthogonality relations on the fourth and fifth subsystems among these four states, the orthogonality graph induced by these four states, after omitting these two subsystems, is identical to case $(1\text{-}3)$ in Fig.~\ref{fig:four-tripartite-222} after an appropriate relabeling of the vertices and a permutation of subsystems. By Lemma~\ref{lem:tripartite-222-exception}, the corresponding tripartite OPSs cannot be perfectly distinguished by LOCC when their three local ranks are all equal to $2$, and can be distinguished by LOCC with nonzero probability otherwise.
	
	\textnormal{(5)} For case $(5\text{-}5)$, the only orthogonality relation on the fifth subsystem is between states $1$ and $5$. Let $\ket{\epsilon_1}$ denote the normalized fifth local component of state $1$. The fifth party performs the measurement $\{M,I-M\}$, where $M=\ket{\epsilon_1}\bra{\epsilon_1}$.
	
	For the outcome corresponding to $M$, state $5$ is excluded, leaving states $\{1,2,3,4\}$. Their vector of the numbers of pairwise orthogonal relations is $(3,1,1,1,0)$. Since there is no orthogonality relation on the fifth subsystem among these four states, their orthogonality relations correspond to the quadripartite vector $(3,1,1,1)$ after omitting the fifth subsystem. By Lemma~\ref{lem:four-ops-multipartite}, these four states can be perfectly distinguished by LOCC.
	
	For the outcome corresponding to $I-M$, state $1$ is excluded, leaving states $\{2,3,4,5\}$. Their vector of the numbers of pairwise orthogonal relations is $(0,2,2,2,0)$. Since there are no orthogonality relations on the first and fifth subsystems among these four states, the orthogonality graph induced by these four states, after omitting these two subsystems, is identical to case $(1\text{-}3)$ in Fig.~\ref{fig:four-tripartite-222} after an appropriate relabeling of the vertices and a permutation of subsystems. By Lemma~\ref{lem:tripartite-222-exception}, the corresponding tripartite OPSs cannot be perfectly distinguished by LOCC when their three local ranks are all equal to $2$, and can be distinguished by LOCC with nonzero probability otherwise.
	
	\textnormal{(6)} For case $(5\text{-}6)$ and $(5\text{-}7)$, the only orthogonality relation on the fifth subsystem is between states $1$ and $5$. Let $\ket{\epsilon_1}$ denote the normalized fifth local component of state $1$. The fifth party performs the measurement $\{M,I-M\}$, where $M=\ket{\epsilon_1}\bra{\epsilon_1}$.
	
	For the outcome corresponding to $M$, state $5$ is excluded, leaving states $\{1,2,3,4\}$. Their vector of the numbers of pairwise orthogonal relations is $(2,2,1,1,0)$. Since there is no orthogonality relation on the fifth subsystem among these four states, their orthogonality relations correspond to the quadripartite vector $(2,2,1,1)$ after omitting the fifth subsystem. By Lemma~\ref{lem:four-ops-multipartite}, these four states can be perfectly distinguished by LOCC.
	
	For the outcome corresponding to $I-M$, state $1$ is excluded, leaving states $\{2,3,4,5\}$. Their vector of the numbers of pairwise orthogonal relations is $(2,0,2,2,0)$. Since there are no orthogonality relations on the second and fifth subsystems among these four states, the orthogonality graph induced by these four states, after omitting these two subsystems, is identical to case $(1\text{-}3)$ in Fig.~\ref{fig:four-tripartite-222} after an appropriate relabeling of the vertices and a permutation of subsystems. By Lemma~\ref{lem:tripartite-222-exception}, the corresponding tripartite OPSs cannot be perfectly distinguished by LOCC when their three local ranks are all equal to $2$, and can be distinguished by LOCC with nonzero probability otherwise.
	
	\textnormal{(7)} For case $(5\text{-}8)$, state $1$ is orthogonal to states $\{2,3\}$ on the first subsystem and to states $\{4,5\}$ on the second subsystem. Since states $4$ and $5$ are orthogonal on the third subsystem rather than on the first subsystem, the conditions of Lemma~\ref{lem:degree-two-two} are satisfied, and state $1$ can be identified by LOCC.
	
	The remaining four states are $\{2,3,4,5\}$. Their vector of the numbers of pairwise orthogonal relations is $(0,0,2,2,2)$. Since there are no orthogonality relations on the first and second subsystems among these four states, the orthogonality graph induced by these four states, after omitting these two subsystems, is identical to case $(1\text{-}3)$ in Fig.~\ref{fig:four-tripartite-222} after an appropriate relabeling of the vertices and a permutation of subsystems. By Lemma~\ref{lem:tripartite-222-exception}, the corresponding tripartite OPSs cannot be perfectly distinguished by LOCC when their three local ranks are all equal to $2$, and can be distinguished by LOCC with nonzero probability otherwise.
	
	We next consider case $(5\text{-}9)$, which corresponds to the second exceptional type in Theorem~\ref{thm:main-distinguishability}. The five OPSs are denoted by $\ket{\psi_1}$, $\ket{\psi_2}$, $\ket{\psi_3}$, $\ket{\psi_4}$, and $\ket{\psi_5}$, where $\ket{\psi_i}$ can be written as $\ket{\alpha_i}_1\ket{\beta_i}_2\ket{\gamma_i}_3 \ket{\delta_i}_4\ket{\epsilon_i}_5$ for $i=1,\ldots,5$.
	
	On the first subsystem, states $1$ and $2$ are orthogonal, and states $3$ and $5$ are orthogonal to each other. Suppose that the first subsystems of states $1$ and $2$ are $\ket{0}$ and $\ket{1}$, respectively.
	
	Since state $5$ is not orthogonal to either state $1$ or state $2$ on the first subsystem, its first subsystem can be written as $\frac{1}{\sqrt{1+|a_1|^2+|b_1|^2}} (\ket{0}+a_1\ket{1}+b_1\ket{2})$, where $a_1\neq0$.
	
	Similarly, state $3$ is not orthogonal to states $1$ and $2$ on the first subsystem. Hence, its first subsystem can be expressed as $\frac{1}{\sqrt{1+|c_1|^2+|d_1|^2+|e_1|^2}} (\ket{0}+c_1\ket{1}+d_1\ket{2}+e_1\ket{3})$, where $c_1\neq0$. Since states $3$ and $5$ are orthogonal on this subsystem, we have $1+a_1^{*}c_1+b_1^{*}d_1=0$.
	
	State $4$ is not orthogonal to any of states $1$, $2$, $3$, or $5$ on the first subsystem. Therefore, its first subsystem can be represented as $\frac{1}{\sqrt{1+|f_1|^2+|g_1|^2+|h_1|^2+|k_1|^2}} (\ket{0}+f_1\ket{1}+g_1\ket{2}+h_1\ket{3}+k_1\ket{4})$, where $f_1\neq0$. Moreover, the absence of additional orthogonality relations gives $1+a_1^{*}f_1+b_1^{*}g_1\neq0$ and $1+c_1^{*}f_1+d_1^{*}g_1+e_1^{*}h_1\neq0$.
	
	Consequently, the general forms of the first subsystems are given by
	\begin{equation}
		\begin{aligned}
			\ket{\alpha_1}_1
			&=\ket{0},\\
			\ket{\alpha_2}_1
			&=\ket{1},\\
			\ket{\alpha_3}_1
			&=
			\frac{
				\ket{0}+c_1\ket{1}+d_1\ket{2}+e_1\ket{3}}
			{\sqrt{1+|c_1|^2+|d_1|^2+|e_1|^2}},\\
			\ket{\alpha_4}_1
			&=
			\frac{
				\ket{0}+f_1\ket{1}+g_1\ket{2}
				+h_1\ket{3}+k_1\ket{4}}
			{\sqrt{1+|f_1|^2+|g_1|^2+|h_1|^2+|k_1|^2}},\\
			\ket{\alpha_5}_1
			&=
			\frac{
				\ket{0}+a_1\ket{1}+b_1\ket{2}}
			{\sqrt{1+|a_1|^2+|b_1|^2}}.
		\end{aligned}
		\label{eq:case82-first}
	\end{equation}
	where $a_1\neq0$, $c_1\neq0$, $f_1\neq0$,
	$1+a_1^{*}c_1+b_1^{*}d_1=0$,
	$1+a_1^{*}f_1+b_1^{*}g_1\neq0$, and
	$1+c_1^{*}f_1+d_1^{*}g_1+e_1^{*}h_1\neq0$; 
	$\frac{\ket{0}+a_1\ket{1}+b_1\ket{2}}{\sqrt{1+|a_1|^2+|b_1|^2}}$ denotes
	$\frac{1}{\sqrt{1+|a_1|^2+|b_1|^2}}(\ket{0}+a_1\ket{1}+b_1\ket{2})$, and similarly for the other expressions.

	On the second subsystem, states $1$ and $3$ are orthogonal, and states $4$ and $5$ are orthogonal to each other. Suppose that the second subsystems of states $1$ and $3$ are $\ket{0}$ and $\ket{1}$, respectively. Following the same construction as that for the first subsystem, the general forms of the second subsystems are
	\begin{equation}
		\begin{aligned}
			\ket{\beta_1}_2
			&=\ket{0},\\
			\ket{\beta_2}_2
			&=
			\frac{
				\ket{0}+f_2\ket{1}+g_2\ket{2}
				+h_2\ket{3}+k_2\ket{4}}
			{\sqrt{1+|f_2|^2+|g_2|^2+|h_2|^2+|k_2|^2}},\\
			\ket{\beta_3}_2
			&=\ket{1},\\
			\ket{\beta_4}_2
			&=
			\frac{
				\ket{0}+a_2\ket{1}+b_2\ket{2}}
			{\sqrt{1+|a_2|^2+|b_2|^2}},\\
			\ket{\beta_5}_2
			&=
			\frac{
				\ket{0}+c_2\ket{1}+d_2\ket{2}+e_2\ket{3}}
			{\sqrt{1+|c_2|^2+|d_2|^2+|e_2|^2}} .
		\end{aligned}
		\label{eq:case82-second}
	\end{equation}
	where $a_2\neq0$, $c_2\neq0$, $f_2\neq0$,
	$1+a_2^{*}c_2+b_2^{*}d_2=0$,
	$1+a_2^{*}f_2+b_2^{*}g_2\neq0$, and
	$1+c_2^{*}f_2+d_2^{*}g_2+e_2^{*}h_2\neq0$.
	
	The constructions on the third, fourth, and fifth subsystems follow the same procedure as those above. On the third subsystem, states $1$ and $4$ are orthogonal, and states $2$ and $3$ are orthogonal. Hence, the corresponding third subsystems can be written as
	\begin{equation}
		\begin{aligned}
			\ket{\gamma_1}_3
			&=\ket{0},\\
			\ket{\gamma_2}_3
			&=
			\frac{
				\ket{0}+a_3\ket{1}+b_3\ket{2}}
			{\sqrt{1+|a_3|^2+|b_3|^2}},\\
			\ket{\gamma_3}_3
			&=
			\frac{
				\ket{0}+c_3\ket{1}+d_3\ket{2}+e_3\ket{3}}
			{\sqrt{1+|c_3|^2+|d_3|^2+|e_3|^2}},\\
			\ket{\gamma_4}_3
			&=\ket{1},\\
			\ket{\gamma_5}_3
			&=
			\frac{
				\ket{0}+f_3\ket{1}+g_3\ket{2}
				+h_3\ket{3}+k_3\ket{4}}
			{\sqrt{1+|f_3|^2+|g_3|^2+|h_3|^2+|k_3|^2}}.
		\end{aligned}
		\label{eq:case82-third}
	\end{equation}
	
	On the fourth subsystem, states $1$ and $5$ are orthogonal, and states $2$ and $4$ are orthogonal. Thus, the general forms of the fourth subsystems are
	\begin{equation}
		\begin{aligned}
			\ket{\delta_1}_4
			&=\ket{0},\\
			\ket{\delta_2}_4
			&=
			\frac{
				\ket{0}+a_4\ket{1}+b_4\ket{2}}
			{\sqrt{1+|a_4|^2+|b_4|^2}},\\
			\ket{\delta_3}_4
			&=
			\frac{
				\ket{0}+f_4\ket{1}+g_4\ket{2}
				+h_4\ket{3}+k_4\ket{4}}
			{\sqrt{1+|f_4|^2+|g_4|^2+|h_4|^2+|k_4|^2}},\\
			\ket{\delta_4}_4
			&=
			\frac{
				\ket{0}+c_4\ket{1}+d_4\ket{2}+e_4\ket{3}}
			{\sqrt{1+|c_4|^2+|d_4|^2+|e_4|^2}},\\
			\ket{\delta_5}_4
			&=\ket{1}.
		\end{aligned}
		\label{eq:case82-fourth}
	\end{equation}
	
	On the fifth subsystem, states $2$ and $5$ are orthogonal, and states $3$ and $4$ are orthogonal. Accordingly, the general forms of the fifth subsystems are
	\begin{equation}
		\begin{aligned}
			\ket{\epsilon_1}_5
			&=
			\frac{
				\ket{0}+f_5\ket{1}+g_5\ket{2}
				+h_5\ket{3}+k_5\ket{4}}
			{\sqrt{1+|f_5|^2+|g_5|^2+|h_5|^2+|k_5|^2}},\\
			\ket{\epsilon_2}_5
			&=\ket{0},\\
			\ket{\epsilon_3}_5
			&=
			\frac{
				\ket{0}+a_5\ket{1}+b_5\ket{2}}
			{\sqrt{1+|a_5|^2+|b_5|^2}},\\
			\ket{\epsilon_4}_5
			&=
			\frac{
				\ket{0}+c_5\ket{1}+d_5\ket{2}+e_5\ket{3}}
			{\sqrt{1+|c_5|^2+|d_5|^2+|e_5|^2}},\\
			\ket{\epsilon_5}_5
			&=\ket{1}.
		\end{aligned}
		\label{eq:case82-fifth}
	\end{equation}
	
	For $p=3,4,5$, the parameters in Eqs.~\eqref{eq:case82-first}--\eqref{eq:case82-fifth} satisfy
	\begin{equation*}
		\begin{aligned}
			&a_p\neq0,\qquad c_p\neq0,\qquad f_p\neq0,\\
			&1+a_p^{*}c_p+b_p^{*}d_p=0,\\
			&1+a_p^{*}f_p+b_p^{*}g_p\neq0,\\
			&1+c_p^{*}f_p+d_p^{*}g_p+e_p^{*}h_p\neq0.
		\end{aligned}
	\end{equation*}
	
	\begin{theorem}
		\label{thm:case82-indistinguishable}
		Five five-partite OPSs corresponding to case $(5\text{-}9)$, with the general forms given in Eqs.~\eqref{eq:case82-first}--\eqref{eq:case82-fifth}, cannot be perfectly distinguished by LOCC when $b_p=d_p=e_p=g_p=h_p=k_p=0$ for $p=1,2,3,4,5$.
	\end{theorem}
	\begin{proof}
		When $b_p=d_p=e_p=g_p=h_p=k_p=0$ for $p=1,2,3,4,5$, the five OPSs corresponding to case $(5\text{-}9)$ take the following form:
		\begingroup\small
		\begin{equation}
			\begin{aligned}
				\ket{\psi_1}
				={}&
				\ket{0}_1
				\ket{0}_2
				\ket{0}_3
				\ket{0}_4
				\frac{\ket{0}+f_5\ket{1}}
				{\sqrt{1+|f_5|^2}}_5,
				\\
				\ket{\psi_2}
				={}&
				\ket{1}_1
				\frac{\ket{0}+f_2\ket{1}}
				{\sqrt{1+|f_2|^2}}_2
				\frac{\ket{0}+a_3\ket{1}}
				{\sqrt{1+|a_3|^2}}_3
				\frac{\ket{0}+a_4\ket{1}}
				{\sqrt{1+|a_4|^2}}_4
				\ket{0}_5,
				\\
				\ket{\psi_3}
				={}&
				\frac{\ket{0}+c_1\ket{1}}
				{\sqrt{1+|c_1|^2}}_1
				\ket{1}_2
				\frac{\ket{0}+c_3\ket{1}}
				{\sqrt{1+|c_3|^2}}_3
				\frac{\ket{0}+f_4\ket{1}}
				{\sqrt{1+|f_4|^2}}_4
				\frac{\ket{0}+a_5\ket{1}}
				{\sqrt{1+|a_5|^2}}_5,
				\\
				\ket{\psi_4}
				={}&
				\frac{\ket{0}+f_1\ket{1}}
				{\sqrt{1+|f_1|^2}}_1
				\frac{\ket{0}+a_2\ket{1}}
				{\sqrt{1+|a_2|^2}}_2
				\ket{1}_3
				\frac{\ket{0}+c_4\ket{1}}
				{\sqrt{1+|c_4|^2}}_4
				\frac{\ket{0}+c_5\ket{1}}
				{\sqrt{1+|c_5|^2}}_5,
				\\
				\ket{\psi_5}
				={}&
				\frac{\ket{0}+a_1\ket{1}}
				{\sqrt{1+|a_1|^2}}_1
				\frac{\ket{0}+c_2\ket{1}}
				{\sqrt{1+|c_2|^2}}_2
				\frac{\ket{0}+f_3\ket{1}}
				{\sqrt{1+|f_3|^2}}_3
				\ket{1}_4
				\ket{1}_5 .
			\end{aligned}
			\label{eq:case82-reduced}
		\end{equation}
		\endgroup
		Here, the orthogonality conditions reduce to $1+a_p^{*}c_p=0$ for $p=1,2,3,4,5$. Under these parameter conditions, the local ranks satisfy $r_1=r_2=r_3=r_4=r_5=2$.
			
		Suppose that the first party performs the first orthogonality-preserving measurement with POVM elements $M_j^{\dagger}M_j$, where $j=1,2,\ldots,l$. Since the local support is two-dimensional, each POVM element can be written in the basis $\{\ket{0},\ket{1}\}$ as
		\[
		M_j^{\dagger}M_j=
		\begin{pmatrix}
			m_{00}^{(j)} & m_{01}^{(j)}\\
			m_{10}^{(j)} & m_{11}^{(j)}
		\end{pmatrix}.
		\]
		
		To preserve the orthogonality of the postmeasurement states, every pair whose orthogonality is provided by the first subsystem must remain orthogonal after the measurement. Since states $\ket{\psi_1}$ and $\ket{\psi_2}$ are orthogonal only on the first subsystem, orthogonality preservation requires $\bra{\psi_1}(M_j^\dagger M_j\otimes I_{2345})\ket{\psi_2}=0$ and $\bra{\psi_2}(M_j^\dagger M_j\otimes I_{2345})\ket{\psi_1}=0$. Hence, $m_{01}^{(j)}=m_{10}^{(j)}=0$.
		
		States $\ket{\psi_3}$ and $\ket{\psi_5}$ are also orthogonal on the first subsystem. Preserving their orthogonality gives $m_{00}^{(j)}+a_1^{*}c_1m_{11}^{(j)}=0$. Since their original orthogonality relation is $1+a_1^{*}c_1=0$, we obtain $m_{00}^{(j)}=m_{11}^{(j)}$. Consequently, $M_j^\dagger M_j=m_{00}^{(j)}I$ on the local support.
		
		If the first party obtains measurement outcome $j$, then $p_1(j|k)= \bra{\psi_k}(M_j^\dagger M_j\otimes I_{2345})\ket{\psi_k} =m_{00}^{(j)}$ for every $k\in\{1,2,3,4,5\}$. Therefore, the outcome probability is the same for all five possible input states, and the first party cannot obtain any useful information.
		
		Suppose next that the second party performs the first orthogonality-preserving measurement. Since states $\ket{\psi_1}$ and $\ket{\psi_3}$ are orthogonal on the second subsystem, the off-diagonal entries of each orthogonality-preserving POVM element must vanish. The orthogonality between states $\ket{\psi_4}$ and $\ket{\psi_5}$, together with $1+a_2^{*}c_2=0$, further implies that the two diagonal entries are equal. Therefore, every orthogonality-preserving POVM element of the second party is proportional to the identity operator on the corresponding local support and provides no information for distinguishing the five states.
		
		For the third party, the orthogonality of states $\ket{\psi_1}$ and $\ket{\psi_4}$ forces the off-diagonal entries of each orthogonality-preserving POVM element to vanish. The orthogonality of states $\ket{\psi_2}$ and $\ket{\psi_3}$, together with $1+a_3^{*}c_3=0$, implies that the two diagonal entries are equal. Hence, every such POVM element is proportional to the identity operator on the local support.
		
		Similarly, for the fourth party, the orthogonality of states $\ket{\psi_1}$ and $\ket{\psi_5}$ forces the off-diagonal entries to vanish, while the orthogonality of states $\ket{\psi_2}$ and $\ket{\psi_4}$, together with $1+a_4^{*}c_4=0$, implies equality of the two diagonal entries. Thus, every orthogonality-preserving POVM element of the fourth party is proportional to the identity operator on the corresponding local support.
		
		For the fifth party, the orthogonality of states $\ket{\psi_2}$ and $\ket{\psi_5}$ forces the off-diagonal entries to vanish. The orthogonality of states $\ket{\psi_3}$ and $\ket{\psi_4}$, together with $1+a_5^{*}c_5=0$, further implies equality of the two diagonal entries. Therefore, every orthogonality-preserving POVM element of the fifth party is also proportional to the identity operator on the corresponding local support.
		
		Consequently, none of the five parties can obtain useful information from an orthogonality-preserving first local measurement. Hence, no nontrivial first local measurement can be performed while preserving the orthogonality of the five states. Therefore, the five five-partite OPSs corresponding to case $(5\text{-}9)$ cannot be perfectly distinguished by LOCC.
	\end{proof}

	\section{Conclusion}
	\label{sec:conclusion}
	
	In this paper, we investigated the local distinguishability of five five-partite OPSs, where any two of the five states are orthogonal on exactly one subsystem. By analyzing the distribution of the ten pairwise orthogonality relations among the five subsystems, we divided these OPSs into seven genuinely five-partite categories.
	
	Up to relabeling of the vertices and permutation of subsystems, these seven categories contain $443$ different orthogonality graphs. We proved that the sets of five OPSs corresponding to $434$ of these graphs can be perfectly distinguished by LOCC. The remaining nine graphs fall into two exceptional structural types, containing eight and one graphs, respectively.
	
	For the $434$ nonexceptional cases, perfect distinguishability was established using orthogonality-preserving local measurements and known results for smaller sets of OPSs. For the first exceptional type, we analyzed the remaining four states after local measurements and examined their orthogonality structures using the known results for four tripartite OPSs. For the second exceptional type, corresponding to case $(5\text{-}9)$, we considered a family of five OPSs satisfying specified parameter conditions. We proved that no party can perform a nontrivial orthogonality-preserving first local measurement on the corresponding local support. Therefore, these five OPSs cannot be perfectly distinguished by LOCC.

	%
	%
	
	\section*{Acknowledgments}
	
	This work is supported by Natural Science Foundation of Shandong Province of China (Grant No. ZR2026MS1078 and ZR2023MF080), Beijing Natural Science Foundation (Grant No. 4252014) and Shandong Key Laboratory of Smart Mine Information Technology.
	%
	

	\appendix
	
	\renewcommand{\thesection}{\Alph{section}}
	\renewcommand{\thesubsection}{\Alph{section}.\arabic{subsection}}
	
	\refstepcounter{section}

	\section*{Appendix~\thesection: Proof of Theorem~\ref{thm:main-distinguishability}}
	\label{app:proof-main}
	The detailed proof of Theorem~\ref{thm:main-distinguishability} is given here.

	\subsection{Category $(6,1,1,1,1)$}
	
	For five five-partite OPSs with the vector of the numbers of pairwise orthogonal relations $(6,1,1,1,1)$, there are six different orthogonality graphs. These graphs are labeled (2-1)--(2-6), as shown in Fig.~\ref{fig:61111}.

	\textnormal{(1)} In cases (2-1) and (2-2), state 1 is orthogonal to all the other states on the first subsystem. Thus, the degree of state 1 on the first subsystem is 4. According to Lemma~\ref{lem:degree-four}, the first party can identify state 1, leaving four states $\{2,3,4,5\}$ to be distinguished. The vector of the numbers of pairwise orthogonal relations among these four states is $(2,1,1,1,1)$, up to a permutation of subsystems. By Lemma~\ref{lem:four-ops-multipartite}, states $\{2,3,4,5\}$ can be perfectly distinguished by LOCC.
	
	\textnormal{(2)} In cases (2-3)--(2-6), state 1 is orthogonal to states $\{2,3,4\}$ on the first subsystem and to state 5 on the second subsystem. Hence, the degree of state 1 is 3 on the first subsystem and 1 on the second subsystem. By Lemma~\ref{lem:degree-three-one}, state 1 can be identified by LOCC. The remaining four states $\{2,3,4,5\}$ have the vector $(3,0,1,1,1)$, up to a permutation of subsystems. Since there is no orthogonality relation on the second subsystem among these four states, their orthogonality relations correspond to the quadripartite vector $(3,1,1,1)$ after omitting the second subsystem. By Lemma~\ref{lem:four-ops-multipartite}, these four states can be perfectly distinguished by LOCC.
	
	Therefore, the sets of five OPSs corresponding to all six cases in the category $(6,1,1,1,1)$ can be perfectly distinguished by LOCC.
	
	\subsection{Category $(5,2,1,1,1)$}
	
	As shown in Fig.~\ref{fig:52111}, there are 30 different orthogonality graphs with the vector $(5,2,1,1,1)$, labeled (3-1)--(3-30).
		
	\textnormal{(1)} In cases $(3\text{-}1)$--$(3\text{-}4)$, state $1$ is orthogonal to all the other states on the first subsystem. By Lemma~\ref{lem:degree-four}, the first party can identify state $1$ by LOCC, leaving states $\{2,3,4,5\}$ to be distinguished. Their vector of the numbers of pairwise orthogonal relations is $(1,2,1,1,1)$. By Lemma~\ref{lem:four-ops-multipartite}, these four states can be perfectly distinguished by LOCC.
	
	\textnormal{(2)} In cases (3-5)--(3-16), state 1 is orthogonal to states $\{2,3,4\}$ on the first subsystem and to state 5 on the second subsystem. Thus, the degree of state 1 is 3 on the first subsystem and 1 on the second subsystem. By Lemma~\ref{lem:degree-three-one}, state 1 can be identified by LOCC. The remaining four states $\{2,3,4,5\}$ have the vector $(2,1,1,1,1)$ and can be perfectly distinguished by Lemma~\ref{lem:four-ops-multipartite}.
	
	\textnormal{(3)} In cases (3-17)--(3-28), state 1 is orthogonal to states $\{2,3,4\}$ on the first subsystem and to state 5 on the third subsystem. Hence, the degree of state 1 is 3 on the first subsystem and 1 on the third subsystem. According to Lemma~\ref{lem:degree-three-one}, state 1 can be identified by LOCC. The remaining states $\{2,3,4,5\}$ have the vector $(2,2,0,1,1)$. Since there is no orthogonality relation among these four states on the third subsystem, they can be regarded as four quadripartite OPSs with the vector $(2,2,1,1)$ and can be perfectly distinguished by Lemma~\ref{lem:four-ops-multipartite}.
	
	\textnormal{(4)} In case (3-29), state 1 is orthogonal to states $\{2,3\}$ on the first subsystem and to states $\{4,5\}$ on the second subsystem. Since states 2 and 3 are not orthogonal on the second subsystem, the conditions of Lemma~\ref{lem:degree-two-two} are satisfied. Thus, the second party can identify state 1 by LOCC. The remaining four states $\{2,3,4,5\}$ have the vector $(3,0,1,1,1)$. Since no two of these states are orthogonal on the second subsystem, they can be regarded as four quadripartite OPSs with the vector $(3,1,1,1)$. By Lemma~\ref{lem:four-ops-multipartite}, these four states can be perfectly distinguished by LOCC.
	
	\textnormal{(5)} In case $(3\text{-}30)$, states $1$ and $5$ are orthogonal on the third subsystem. The third party can perform a local measurement such that the remaining possible states are either $\{1,2,3,4\}$ or $\{2,3,4,5\}$.
	
	For states $\{1,2,3,4\}$, the vector of the numbers of pairwise orthogonal relations is $(3,2,0,0,1)$. Since there is no orthogonality relation on the third and fourth subsystems, these states can be regarded as four tripartite OPSs with the vector $(3,2,1)$. By Lemma~\ref{lem:tripartite-411-321}, states $\{1,2,3,4\}$ can be perfectly distinguished by LOCC.
	
	For states $\{2,3,4,5\}$, the vector of the numbers of pairwise orthogonal relations is $(3,1,0,1,1)$. Omitting the third subsystem gives four quadripartite OPSs with the vector $(3,1,1,1)$, which can be perfectly distinguished by Lemma~\ref{lem:four-ops-multipartite}.
	
	\subsection{Category $(4,3,1,1,1)$}
	
	For the vector $(4,3,1,1,1)$, there are 47 different orthogonality graphs, labeled (4-1)--(4-47), as shown in Fig.~\ref{fig:43111}.
	
	\textnormal{(1)} In cases $(4\text{-}1)$--$(4\text{-}3)$, state $1$ is orthogonal to all the other states on the first subsystem. By Lemma~\ref{lem:degree-four}, the first party can identify state $1$ by LOCC, leaving states $\{2,3,4,5\}$ to be distinguished. Their vector of the numbers of pairwise orthogonal relations is $(0,3,1,1,1)$. Since there is no orthogonality relation on the first subsystem among these four states, their orthogonality relations correspond to the quadripartite vector $(3,1,1,1)$ after omitting the first subsystem. By Lemma~\ref{lem:four-ops-multipartite}, these four states can be perfectly distinguished by LOCC.
	
	\textnormal{(2)} In cases (4-4)--(4-15), state 1 is orthogonal to states $\{2,3,4\}$ on the first subsystem and to state 5 on the second subsystem. Hence, the degree of state 1 is 3 on the first subsystem and 1 on the second subsystem. By Lemma~\ref{lem:degree-three-one}, state 1 can be identified by LOCC. The remaining four states $\{2,3,4,5\}$ have the vector $(1,2,1,1,1)$ and can be perfectly distinguished by Lemma~\ref{lem:four-ops-multipartite}.
	
	\textnormal{(3)} In cases (4-16)--(4-27), state 1 is orthogonal to states $\{2,3,4\}$ on the first subsystem and to state 5 on the third subsystem. Thus, the degree of state 1 is 3 on the first subsystem and 1 on the third subsystem. According to Lemma~\ref{lem:degree-three-one}, state 1 can be identified by LOCC. The remaining states $\{2,3,4,5\}$ have the vector $(1,3,0,1,1)$. Since there is no orthogonality relation among these four states on the third subsystem, they can be regarded as four quadripartite OPSs with the vector $(1,3,1,1)$ and can be perfectly distinguished by Lemma~\ref{lem:four-ops-multipartite}.
	
	\textnormal{(4)} In cases (4-28), (4-34), (4-36), and (4-38), state 1 is orthogonal to states $\{2,3\}$ on the first subsystem and to states $\{4,5\}$ on the second subsystem. Since states 2 and 3 are not orthogonal on the second subsystem, the conditions of Lemma~\ref{lem:degree-two-two} are satisfied. Thus, the second party can identify state 1 by LOCC. The remaining four states $\{2,3,4,5\}$ have the vector $(2,1,1,1,1)$ and can be perfectly distinguished by Lemma~\ref{lem:four-ops-multipartite}.
	
	\textnormal{(5)} In cases (4-29), (4-30), (4-32), (4-33), (4-35), and (4-37), state 1 is orthogonal to states $\{2,3\}$ on the first subsystem and to states $\{4,5\}$ on the second subsystem. Since states 4 and 5 are not orthogonal on the first subsystem, the conditions of Lemma~\ref{lem:degree-two-two} are satisfied. Therefore, the first party can identify state 1. The remaining four states $\{2,3,4,5\}$ have the vector $(2,1,1,1,1)$ and can be perfectly distinguished by Lemma~\ref{lem:four-ops-multipartite}.
	
	\textnormal{(6)} In case $(4\text{-}31)$, states $2$ and $5$ are orthogonal on the third subsystem. The third party can perform a local measurement such that the remaining possible states are either $\{1,2,3,4\}$ or $\{1,3,4,5\}$.
	
	For states $\{1,2,3,4\}$, the vector of the numbers of pairwise orthogonal relations is $(3,2,0,1,0)$. Since there is no orthogonality relation on the third and fifth subsystems, these states can be regarded as four tripartite OPSs with the vector $(3,2,1)$. By Lemma~\ref{lem:tripartite-411-321}, states $\{1,2,3,4\}$ can be perfectly distinguished by LOCC.
	
	For states $\{1,3,4,5\}$, the vector is $(2,2,0,1,1)$. Omitting the third subsystem gives four quadripartite OPSs with the vector $(2,2,1,1)$, which can be perfectly distinguished by Lemma~\ref{lem:four-ops-multipartite}.
	
	\textnormal{(7)} In case (4-39), the degree of state 4 is 3 on the second subsystem and 1 on the first subsystem. According to Lemma~\ref{lem:degree-three-one}, the second party can identify state 4 by LOCC. The remaining four states $\{1,2,3,5\}$ have the vector $(3,0,1,1,1)$. Since no two of these states are orthogonal on the second subsystem, they can be regarded as four quadripartite OPSs with the vector $(3,1,1,1)$ and can be perfectly distinguished by Lemma~\ref{lem:four-ops-multipartite}.
	
	\textnormal{(8)} In case $(4\text{-}40)$, states $1$ and $5$ are orthogonal on the third subsystem. The third party can perform a local measurement such that the remaining possible states are either $\{1,2,3,4\}$ or $\{2,3,4,5\}$.
	
	For states $\{1,2,3,4\}$, the vector of the numbers of pairwise orthogonal relations is $(3,2,0,0,1)$. Omitting the third and fourth subsystems gives four tripartite OPSs with the vector $(3,2,1)$, which can be perfectly distinguished by Lemma~\ref{lem:tripartite-411-321}.
	
	For states $\{2,3,4,5\}$, the vector is $(2,2,0,1,1)$. Since there is no orthogonality relation on the third subsystem, these states can be regarded as four quadripartite OPSs with the vector $(2,2,1,1)$ and can be perfectly distinguished by Lemma~\ref{lem:four-ops-multipartite}.
	
	\textnormal{(9)} In cases (4-41), (4-43), (4-45), and (4-46), the degree of state 1 is 2 on the first subsystem, 1 on the second subsystem, and 1 on the third subsystem. Since states 4 and 5 are not orthogonal on the first subsystem, the conditions of Lemma~\ref{lem:degree-two-one-one} are satisfied. Thus, the first party can identify state 1 by LOCC. The remaining four states $\{2,3,4,5\}$ have the vector $(2,2,0,1,1)$. Omitting the third subsystem, these states can be regarded as four quadripartite OPSs with the vector $(2,2,1,1)$. By Lemma~\ref{lem:four-ops-multipartite}, they can be perfectly distinguished by LOCC.
	
	\textnormal{(10)} In case (4-42), the degree of state 2 is 2 on the second subsystem, 1 on the first subsystem, and 1 on the fourth subsystem. Since states 1 and 5 are not orthogonal on the second subsystem, Lemma~\ref{lem:degree-two-one-one} implies that the second party can identify state 2 by LOCC. The remaining four states $\{1,3,4,5\}$ have the vector $(3,1,1,0,1)$. Since there is no orthogonality relation on the fourth subsystem, they can be regarded as four quadripartite OPSs with the vector $(3,1,1,1)$ and can be perfectly distinguished by Lemma~\ref{lem:four-ops-multipartite}.
	
	\textnormal{(11)} In case (4-44), the degree of state 2 is 2 on the first subsystem, 1 on the second subsystem, and 1 on the fourth subsystem. Since states 3 and 4 are not orthogonal on the first subsystem, the conditions of Lemma~\ref{lem:degree-two-one-one} are satisfied. Thus, the first party can identify state 2 by LOCC. The remaining four states $\{1,3,4,5\}$ have the vector $(2,2,1,0,1)$. Omitting the fourth subsystem gives four quadripartite OPSs with the vector $(2,2,1,1)$, which can be perfectly distinguished by Lemma~\ref{lem:four-ops-multipartite}.
	
	\textnormal{(12)} In case (4-47), the degree of state 2 is 2 on the second subsystem, 1 on the first subsystem, and 1 on the fourth subsystem. Since states 1 and 3 are not orthogonal on the second subsystem, the conditions of Lemma~\ref{lem:degree-two-one-one} are satisfied. According to this lemma, the second party can identify state 2 by LOCC. The remaining four states $\{1,3,4,5\}$ have the vector $(3,1,1,0,1)$. After omitting the fourth subsystem, these four quadripartite OPSs have the vector $(3,1,1,1)$ and can be perfectly distinguished by Lemma~\ref{lem:four-ops-multipartite}.

	\subsection{Category $(4,2,2,1,1)$}

	For the category $(4,2,2,1,1)$, the sets of five OPSs corresponding to 96 orthogonality graphs can be perfectly distinguished by LOCC. These graphs are labeled $(6\text{-}1)$--$(6\text{-}96)$, as shown in Fig.~\ref{fig:42211-2}.
	
	\textnormal{(1)} In cases (6-1)--(6-4), state 1 is orthogonal to all the other states on the first subsystem. According to Lemma~\ref{lem:degree-four}, the first party can identify state 1. The remaining four states $\{2,3,4,5\}$ have the vector $(0,2,2,1,1)$. Omitting the first subsystem gives four quadripartite OPSs with the vector $(2,2,1,1)$, which can be perfectly distinguished by Lemma~\ref{lem:four-ops-multipartite}.
	
	\textnormal{(2)} In cases (6-5)--(6-36), state 1 is orthogonal to states $\{2,3,4\}$ on the first subsystem and to state 5 on the second subsystem. According to Lemma~\ref{lem:degree-three-one}, state 1 can be identified by LOCC. The remaining states $\{2,3,4,5\}$ have the vector $(1,1,2,1,1)$ and can be perfectly distinguished by Lemma~\ref{lem:four-ops-multipartite}.
	
	\textnormal{(3)} In cases (6-37)--(6-54), state 1 is orthogonal to states $\{2,3,4\}$ on the first subsystem and to state 5 on the fourth subsystem. By Lemma~\ref{lem:degree-three-one}, state 1 can be identified by LOCC. The remaining four states $\{2,3,4,5\}$ have the vector $(1,2,2,0,1)$. Since there is no orthogonality relation on the fourth subsystem, they can be regarded as four quadripartite OPSs with the vector $(1,2,2,1)$ and can be perfectly distinguished by Lemma~\ref{lem:four-ops-multipartite}.
	
	\textnormal{(4)} In cases (6-55)--(6-64), state 1 is orthogonal to states $\{2,3\}$ on the first subsystem and to states $\{4,5\}$ on the second subsystem. Since states 2 and 3 are not orthogonal on the second subsystem, the second party can identify state 1 by Lemma~\ref{lem:degree-two-two}. The remaining states $\{2,3,4,5\}$ have the vector $(2,0,2,1,1)$, which reduces to the quadripartite vector $(2,2,1,1)$. By Lemma~\ref{lem:four-ops-multipartite}, these four states can be perfectly distinguished by LOCC.
	
	\textnormal{(5)} In case (6-65), state 4 is orthogonal to states $\{1,2\}$ on the second subsystem, to state 5 on the first subsystem, and to state 3 on the fourth subsystem. Since states $3$ and $5$ are not orthogonal on the second subsystem, the conditions of Lemma~\ref{lem:degree-two-one-one} are satisfied. By Lemma~\ref{lem:degree-two-one-one}, the second party can identify state 4. The remaining states $\{1,2,3,5\}$ have the vector $(3,0,2,0,1)$, which reduces to the tripartite vector $(3,2,1)$. These four states can be perfectly distinguished by Lemma~\ref{lem:tripartite-411-321}.
	
	\textnormal{(6)} In case (6-66), state 4 is orthogonal to states $\{1,2\}$ on the second subsystem, to state 5 on the first subsystem, and to state 3 on the third subsystem. Since states $3$ and $5$ are not orthogonal on the second subsystem, the conditions of Lemma~\ref{lem:degree-two-one-one} are satisfied. According to Lemma~\ref{lem:degree-two-one-one}, the second party can identify state 4. The remaining four states $\{1,2,3,5\}$ have the vector $(3,0,1,1,1)$. Omitting the second subsystem gives four quadripartite OPSs with the vector $(3,1,1,1)$, which can be perfectly distinguished by Lemma~\ref{lem:four-ops-multipartite}.
	
	\textnormal{(7)} For case (6-67), state 4 is orthogonal to states $\{1,2\}$ on the second subsystem, to state 5 on the first subsystem, and to state 3 on the fifth subsystem. Since states $3$ and $5$ are not orthogonal on the second subsystem, the conditions of Lemma~\ref{lem:degree-two-one-one} are satisfied. By Lemma~\ref{lem:degree-two-one-one}, state 4 can be identified by the second party. The remaining states $\{1,2,3,5\}$ have the vector $(3,0,2,1,0)$. Omitting the second and fifth subsystems gives four tripartite OPSs with the vector $(3,2,1)$, which can be perfectly distinguished by Lemma~\ref{lem:tripartite-411-321}.
	
	\textnormal{(8)} In cases (6-68), (6-69), (6-70), (6-72), and (6-73), state 1 is orthogonal to states $\{2,3\}$ on the first subsystem, to state 4 on the second subsystem, and to state 5 on the third subsystem. Since states 4 and 5 are not orthogonal on the first subsystem, the first party can identify state 1 by Lemma~\ref{lem:degree-two-one-one}. The remaining states $\{2,3,4,5\}$ have the vector $(2,1,1,1,1)$ and can be perfectly distinguished by Lemma~\ref{lem:four-ops-multipartite}.
	
	\textnormal{(9)} In case (6-71), state 2 is orthogonal to states $\{1,4\}$ on the first subsystem, to state 3 on the second subsystem, and to state 5 on the fourth subsystem. Since states 3 and 5 are not orthogonal on the first subsystem, state 2 can be identified by the first party according to Lemma~\ref{lem:degree-two-one-one}. The remaining four states $\{1,3,4,5\}$ have the vector $(2,1,2,0,1)$, which reduces to the quadripartite vector $(2,1,2,1)$. They can be perfectly distinguished by Lemma~\ref{lem:four-ops-multipartite}.
	
	\textnormal{(10)} In case (6-74), state 3 is orthogonal to states $\{1,5\}$ on the first subsystem, to state 2 on the second subsystem, and to state 4 on the fifth subsystem. By Lemma~\ref{lem:degree-two-one-one}, the first party can identify state 3. The remaining states $\{1,2,4,5\}$ have the vector $(2,1,2,1,0)$. Since there is no orthogonality relation on the fifth subsystem, they can be regarded as four quadripartite OPSs with the vector $(2,1,2,1)$ and can be perfectly distinguished by Lemma~\ref{lem:four-ops-multipartite}.
	
	\textnormal{(11)} In cases (6-75)--(6-79), state 1 is orthogonal to states $\{2,3\}$ on the first subsystem, to state 4 on the second subsystem, and to state 5 on the fourth subsystem. According to Lemma~\ref{lem:degree-two-one-one}, the first party can identify state 1. The remaining four states $\{2,3,4,5\}$ have the vector $(2,1,2,0,1)$, which reduces to the quadripartite vector $(2,1,2,1)$. By Lemma~\ref{lem:four-ops-multipartite}, these states can be perfectly distinguished by LOCC.
	
	\textnormal{(12)} In case (6-80), state 2 is orthogonal to states $\{1,5\}$ on the first subsystem, to state 3 on the second subsystem, and to state 4 on the third subsystem. By Lemma~\ref{lem:degree-two-one-one}, the first party can identify state 2. The remaining states $\{1,3,4,5\}$ have the vector $(2,1,1,1,1)$ and can be perfectly distinguished by Lemma~\ref{lem:four-ops-multipartite}.
	
	\textnormal{(13)} In case (6-81), states $1$ and $4$ are orthogonal on the fourth subsystem. The fourth party can perform a local measurement such that the remaining possible states are either $\{1,2,3,5\}$ or $\{2,3,4,5\}$.
	
	For states $\{1,2,3,5\}$, the vector of the numbers of pairwise orthogonal relations is $(3,1,1,0,1)$. Omitting the fourth subsystem gives four quadripartite OPSs with the vector $(3,1,1,1)$, which can be perfectly distinguished by Lemma~\ref{lem:four-ops-multipartite}.
	
	For states $\{2,3,4,5\}$, the vector is $(2,1,2,0,1)$, corresponding to the quadripartite vector $(2,1,2,1)$. By Lemma~\ref{lem:four-ops-multipartite}, these four states can also be perfectly distinguished by LOCC.
	
	\textnormal{(14)} In case (6-82), state 2 is orthogonal to states $\{1,4\}$ on the first subsystem, to state 3 on the third subsystem, and to state 5 on the fifth subsystem. According to Lemma~\ref{lem:degree-two-one-one}, the first party can identify state 2. The remaining states $\{1,3,4,5\}$ have the vector $(2,2,1,1,0)$. Omitting the fifth subsystem gives four quadripartite OPSs with the vector $(2,2,1,1)$, which can be perfectly distinguished by Lemma~\ref{lem:four-ops-multipartite}.
	
	\textnormal{(15)} In cases (6-83)--(6-88), state 1 is orthogonal to states $\{4,5\}$ on the first subsystem and to states $\{2,3\}$ on the second subsystem. Since states 4 and 5 are not orthogonal on the second subsystem, the second party can identify state 1 by Lemma~\ref{lem:degree-two-two}. The remaining four states $\{2,3,4,5\}$ have the vector $(2,0,2,1,1)$, which reduces to the quadripartite vector $(2,2,1,1)$. Their perfect distinguishability follows from Lemma~\ref{lem:four-ops-multipartite}.
	
	\textnormal{(16)} In cases (6-89)--(6-92), state 1 is orthogonal to states $\{2,3\}$ on the second subsystem, to state 4 on the first subsystem, and to state 5 on the third subsystem. By Lemma~\ref{lem:degree-two-one-one}, the second party can identify state 1. The remaining states $\{2,3,4,5\}$ have the vector $(3,0,1,1,1)$. Since there is no orthogonality relation on the second subsystem, they can be regarded as four quadripartite OPSs with the vector $(3,1,1,1)$ and can be perfectly distinguished by Lemma~\ref{lem:four-ops-multipartite}.
	
	\textnormal{(17)} In case (6-93), state 1 is orthogonal to states $\{2,3\}$ on the second subsystem, to state 4 on the first subsystem, and to state 5 on the fourth subsystem. According to Lemma~\ref{lem:degree-two-one-one}, the second party can identify state 1. The remaining states $\{2,3,4,5\}$ have the vector $(3,0,2,0,1)$, which reduces to the tripartite vector $(3,2,1)$. By Lemma~\ref{lem:tripartite-411-321}, these four states can be perfectly distinguished by LOCC.
	
	\textnormal{(18)} For case $(6\text{-}94)$, state $1$ is orthogonal to states $\{2,3\}$ on the second subsystem, to state $5$ on the first subsystem, and to state $4$ on the third subsystem. By Lemma~\ref{lem:degree-two-one-one}, the second party can identify state $1$ by LOCC. The remaining four states $\{2,3,4,5\}$ have the vector $(3,0,1,1,1)$. Since there is no orthogonality relation on the second subsystem among these four states, their orthogonality relations correspond to the quadripartite vector $(3,1,1,1)$ after omitting the second subsystem. By Lemma~\ref{lem:four-ops-multipartite}, these four states can be perfectly distinguished by LOCC.
	
	\textnormal{(19)} In cases (6-95) and (6-96), state 1 is orthogonal to states $\{2,3\}$ on the second subsystem and to states $\{4,5\}$ on the third subsystem. Since states 4 and 5 are not orthogonal on the second subsystem, the second party can identify state 1 by Lemma~\ref{lem:degree-two-two}. The remaining four states $\{2,3,4,5\}$ have the vector $(4,0,0,1,1)$. Omitting the second and third subsystems gives four tripartite OPSs with the vector $(4,1,1)$, which can be perfectly distinguished by Lemma~\ref{lem:tripartite-411-321}.
	
	Therefore, the sets of five OPSs corresponding to all 96 cases in the category $(4,2,2,1,1)$ can be perfectly distinguished by LOCC.
	
	\subsection{Category $(3,3,2,1,1)$}

	For the category $(3,3,2,1,1)$, the sets of five OPSs corresponding to 120 orthogonality graphs can be perfectly distinguished by LOCC. These graphs are labeled $(7\text{-}1)$--$(7\text{-}120)$, as shown in Fig.~\ref{fig:33211-2}.
	
	\textnormal{(1)} In cases (7-1)--(7-15), state 1 is orthogonal to states $\{2,3,4\}$ on the first subsystem and to state 5 on the second subsystem. According to Lemma~\ref{lem:degree-three-one}, state 1 can be identified by LOCC. The remaining four states $\{2,3,4,5\}$ have the vector $(0,2,2,1,1)$. Omitting the first subsystem gives four quadripartite OPSs with the vector $(2,2,1,1)$, which can be perfectly distinguished by Lemma~\ref{lem:four-ops-multipartite}.
	
	\textnormal{(2)} In cases (7-16)--(7-27), state 1 is orthogonal to states $\{2,3,4\}$ on the first subsystem and to state 5 on the third subsystem. By Lemma~\ref{lem:degree-three-one}, state 1 can be identified by LOCC. The remaining states $\{2,3,4,5\}$ have the vector $(0,3,1,1,1)$, which reduces to the quadripartite vector $(3,1,1,1)$. These four states can be perfectly distinguished by Lemma~\ref{lem:four-ops-multipartite}.
	
	\textnormal{(3)} For cases (7-28)--(7-38), state 1 is orthogonal to states $\{2,3,4\}$ on the first subsystem and to state 5 on the fourth subsystem. According to Lemma~\ref{lem:degree-three-one}, state 1 can be identified by LOCC. The remaining four states $\{2,3,4,5\}$ have the vector $(0,3,2,0,1)$. Since there is no orthogonality relation on the first and fourth subsystems, these states can be regarded as four tripartite OPSs with the vector $(3,2,1)$ and can be perfectly distinguished by Lemma~\ref{lem:tripartite-411-321}.
	
	\textnormal{(4)} In cases (7-39)--(7-52) and (7-55)--(7-64), state 1 is orthogonal to states $\{2,3\}$ on the first subsystem and to states $\{4,5\}$ on the second subsystem. Since states 4 and 5 are not orthogonal on the first subsystem, the first party can identify state 1 by Lemma~\ref{lem:degree-two-two}. The remaining states $\{2,3,4,5\}$ have the vector $(1,1,2,1,1)$ and can be perfectly distinguished by Lemma~\ref{lem:four-ops-multipartite}.
	
	\textnormal{(5)} In case (7-53), state 2 is orthogonal to states $\{4,5\}$ on the third subsystem, to state 1 on the first subsystem, and to state 3 on the second subsystem. By Lemma~\ref{lem:degree-two-one-one}, the third party can identify state 2. The remaining four states $\{1,3,4,5\}$ have the vector $(2,2,0,1,1)$. Omitting the third subsystem gives four quadripartite OPSs with the vector $(2,2,1,1)$, which can be perfectly distinguished by Lemma~\ref{lem:four-ops-multipartite}.
	
	\textnormal{(6)} In case (7-54), state 2 is orthogonal to state 4 on the fourth subsystem, which can be used to distinguish the sets $\{1,2,3,5\}$ and $\{1,3,4,5\}$ by the fourth party. For both sets, the vector of the numbers of pairwise orthogonal relations is $(2,2,1,0,1)$. Omitting the fourth subsystem gives four quadripartite OPSs with the vector $(2,2,1,1)$. By Lemma~\ref{lem:four-ops-multipartite}, both sets can be perfectly distinguished by LOCC.
	
	\textnormal{(7)} In cases (7-65)--(7-74), (7-76), (7-78), and (7-80), state 1 is orthogonal to states $\{2,3\}$ on the first subsystem, to state 4 on the second subsystem, and to state 5 on the third subsystem. Since states 4 and 5 are not orthogonal on the first subsystem, the first party can identify state 1 by Lemma~\ref{lem:degree-two-one-one}. The remaining four states $\{2,3,4,5\}$ have the vector $(1,2,1,1,1)$ and can be perfectly distinguished by Lemma~\ref{lem:four-ops-multipartite}.
	
	\textnormal{(8)} In cases (7-75), (7-84), and (7-85)--(7-87), state 2 is orthogonal to states $\{3,4\}$ on the second subsystem, to state 1 on the first subsystem, and to state 5 on the third subsystem. According to Lemma~\ref{lem:degree-two-one-one}, the second party can identify state 2. The remaining four states $\{1,3,4,5\}$ have the vector $(2,1,1,1,1)$ and can be perfectly distinguished by Lemma~\ref{lem:four-ops-multipartite}.
	
	\textnormal{(9)} In cases (7-77) and (7-79), state 2 is orthogonal to states $\{3,4\}$ on the second subsystem, to state 1 on the first subsystem, and to state 5 on the fourth subsystem. By Lemma~\ref{lem:degree-two-one-one}, the second party can identify state 2. The remaining states $\{1,3,4,5\}$ have the vector $(2,1,2,0,1)$. Omitting the fourth subsystem gives four quadripartite OPSs with the vector $(2,1,2,1)$, which can be perfectly distinguished by Lemma~\ref{lem:four-ops-multipartite}.
	
	\textnormal{(10)} In case (7-81), state 2 is orthogonal to state 5 on the fourth subsystem, which can be used to distinguish the sets $\{1,2,3,4\}$ and $\{1,3,4,5\}$ by the fourth party. The vector of the numbers of pairwise orthogonal relations for both sets is $(2,2,1,0,1)$. Hence, both sets can be regarded as four quadripartite OPSs with the vector $(2,2,1,1)$ and can be perfectly distinguished by Lemma~\ref{lem:four-ops-multipartite}.
	
	\textnormal{(11)} For cases (7-82), (7-83), and (7-89), state 1 is orthogonal to states $\{2,3\}$ on the first subsystem, to state 4 on the second subsystem, and to state 5 on the fourth subsystem. According to Lemma~\ref{lem:degree-two-one-one}, the first party can identify state 1. The remaining four states $\{2,3,4,5\}$ have the vector $(1,2,2,0,1)$, which reduces to the quadripartite vector $(1,2,2,1)$. They can be perfectly distinguished by Lemma~\ref{lem:four-ops-multipartite}.
	
	\textnormal{(12)} In case (7-88), state 2 is orthogonal to states $\{3,4\}$ on the second subsystem, to state 1 on the first subsystem, and to state 5 on the fifth subsystem. By Lemma~\ref{lem:degree-two-one-one}, the second party can identify state 2. The remaining states $\{1,3,4,5\}$ have the vector $(2,1,2,1,0)$. Since there is no orthogonality relation on the fifth subsystem, they can be regarded as four quadripartite OPSs with the vector $(2,1,2,1)$ and can be perfectly distinguished by Lemma~\ref{lem:four-ops-multipartite}.
	
	\textnormal{(13)} In cases (7-90)--(7-94), (7-96), and (7-98), state 1 is orthogonal to states $\{2,3\}$ on the first subsystem, to state 5 on the second subsystem, and to state 4 on the third subsystem. According to Lemma~\ref{lem:degree-two-one-one}, the first party can identify state 1. The remaining four states $\{2,3,4,5\}$ have the vector $(1,2,1,1,1)$ and can be perfectly distinguished by Lemma~\ref{lem:four-ops-multipartite}.
	
	\textnormal{(14)} In case (7-95), state 2 is orthogonal to state 5 on the fourth subsystem, which can be used to distinguish the sets $\{1,2,3,4\}$ and $\{1,3,4,5\}$ by the fourth party.
	
	For states $\{1,2,3,4\}$, the vector of the numbers of pairwise orthogonal relations is $(2,2,2,0,0)$. Since there are no orthogonality relations on the fourth and fifth subsystems among these four states, the orthogonality graph induced by these states is identical to case $(1\text{-}1)$ in Fig.~\ref{fig:four-tripartite-222} after an appropriate relabeling of the vertices and a permutation of subsystems. By Lemma~\ref{lem:tripartite-222-dist}, these four states can be perfectly distinguished by LOCC.
	
	For states $\{1,3,4,5\}$, the vector is $(2,2,1,0,1)$, which reduces to the quadripartite vector $(2,2,1,1)$. By Lemma~\ref{lem:four-ops-multipartite}, these four states can also be perfectly distinguished by LOCC.
	
	\textnormal{(15)} In case (7-97), state 2 is orthogonal to state 5 on the fourth subsystem, allowing the fourth party to distinguish the sets $\{1,2,3,4\}$ and $\{1,3,4,5\}$.
	
	For states $\{1,2,3,4\}$, the vector is $(2,1,2,0,1)$, which reduces to the quadripartite vector $(2,1,2,1)$. For states $\{1,3,4,5\}$, the vector is $(2,2,1,0,1)$, corresponding to the quadripartite vector $(2,2,1,1)$. Both sets can be perfectly distinguished by Lemma~\ref{lem:four-ops-multipartite}.
	
	\textnormal{(16)} In cases (7-99)--(7-104), state 1 is orthogonal to states $\{2,3\}$ on the first subsystem and to states $\{4,5\}$ on the third subsystem. Since states 4 and 5 are not orthogonal on the first subsystem, the first party can identify state 1 by Lemma~\ref{lem:degree-two-two}. The remaining states $\{2,3,4,5\}$ have the vector $(1,3,0,1,1)$. Omitting the third subsystem gives four quadripartite OPSs with the vector $(1,3,1,1)$, which can be perfectly distinguished by Lemma~\ref{lem:four-ops-multipartite}.
	
	\textnormal{(17)} In cases (7-105) and (7-106), state 1 is orthogonal to states $\{2,3\}$ on the first subsystem, to state 4 on the third subsystem, and to state 5 on the fourth subsystem. By Lemma~\ref{lem:degree-two-one-one}, the first party can identify state 1. The remaining four states $\{2,3,4,5\}$ have the vector $(1,3,1,0,1)$, which reduces to the quadripartite vector $(1,3,1,1)$. Their perfect distinguishability follows from Lemma~\ref{lem:four-ops-multipartite}.
	
	\textnormal{(18)} For cases (7-107) and (7-108), state 1 is orthogonal to states $\{2,3\}$ on the first subsystem, to state 5 on the second subsystem, and to state 4 on the fourth subsystem. According to Lemma~\ref{lem:degree-two-one-one}, the first party can identify state 1. The remaining states $\{2,3,4,5\}$ have the vector $(1,2,2,0,1)$, which reduces to the quadripartite vector $(1,2,2,1)$. They can be perfectly distinguished by Lemma~\ref{lem:four-ops-multipartite}.
	
	\textnormal{(19)} In cases (7-109)--(7-111), state 1 is orthogonal to states $\{4,5\}$ on the first subsystem and to states $\{2,3\}$ on the third subsystem. Since states 4 and 5 are not orthogonal on the third subsystem, the third party can identify state 1 by Lemma~\ref{lem:degree-two-two}. The remaining four states $\{2,3,4,5\}$ have the vector $(1,3,0,1,1)$, which reduces to the quadripartite vector $(1,3,1,1)$. By Lemma~\ref{lem:four-ops-multipartite}, these four states can be perfectly distinguished by LOCC.
	
	\textnormal{(20)} Finally, in cases (7-112)--(7-120), state 1 is orthogonal to states $\{2,3\}$ on the third subsystem, to state 4 on the first subsystem, and to state 5 on the second subsystem. By Lemma~\ref{lem:degree-two-one-one}, the third party can identify state 1. The remaining states $\{2,3,4,5\}$ have the vector $(2,2,0,1,1)$. Omitting the third subsystem gives four quadripartite OPSs with the vector $(2,2,1,1)$, which can be perfectly distinguished by Lemma~\ref{lem:four-ops-multipartite}.
	
	Therefore, the 120 cases labeled (7-1)--(7-120) can be perfectly distinguished by LOCC.

	\subsection{Category $(3,2,2,2,1)$}
	
	For the category $(3,2,2,2,1)$, the sets of five OPSs corresponding to 122 orthogonality graphs can be perfectly distinguished by LOCC. These graphs are labeled $(8\text{-}1)$--$(8\text{-}122)$, as shown in Fig.~\ref{fig:32221-2}.
	
	\textnormal{(1)} In cases (8-1)--(8-18), state 1 is orthogonal to states $\{2,3,4\}$ on the first subsystem and to state 5 on the second subsystem. According to Lemma~\ref{lem:degree-three-one}, the first party can identify state 1. The remaining four states $\{2,3,4,5\}$ have the vector $(0,1,2,2,1)$. Omitting the first subsystem gives four quadripartite OPSs with the vector $(1,2,2,1)$, which can be perfectly distinguished by Lemma~\ref{lem:four-ops-multipartite}.
	
	\textnormal{(2)} In cases (8-19)--(8-22), state 1 is orthogonal to states $\{2,3,4\}$ on the first subsystem and to state 5 on the fifth subsystem. By Lemma~\ref{lem:degree-three-one}, state 1 can be identified by LOCC. The remaining states $\{2,3,4,5\}$ have the vector $(0,2,2,2,0)$. After omitting the first and fifth subsystems, their orthogonality graphs correspond to cases $(1\text{-}1)$ or $(1\text{-}2)$ in Fig.~\ref{fig:four-tripartite-222}, up to relabeling of the vertices and permutation of subsystems. By Lemma~\ref{lem:tripartite-222-dist}, these four states can be perfectly distinguished by LOCC.
	
	\textnormal{(3)} In cases (8-23)--(8-42), state 1 is orthogonal to states $\{2,3\}$ on the first subsystem and to states $\{4,5\}$ on the second subsystem. Since states 2 and 3 are not orthogonal on the second subsystem, the second party can identify state 1 by Lemma~\ref{lem:degree-two-two}. The remaining states $\{2,3,4,5\}$ have the vector $(1,0,2,2,1)$, which reduces to the quadripartite vector $(1,2,2,1)$. They can be perfectly distinguished by Lemma~\ref{lem:four-ops-multipartite}.
	
	\textnormal{(4)} In cases (8-43)--(8-56) and (8-59)--(8-66), state 1 is orthogonal to states $\{2,3\}$ on the first subsystem, to state 4 on the second subsystem, and to state 5 on the third subsystem. Since states 4 and 5 are not orthogonal on the first subsystem, the first party can identify state 1 according to Lemma~\ref{lem:degree-two-one-one}. The remaining four states $\{2,3,4,5\}$ have the vector $(1,1,1,2,1)$ and can be perfectly distinguished by Lemma~\ref{lem:four-ops-multipartite}.
	
	\textnormal{(5)} In case (8-57), state 3 is orthogonal to state 5 on the fifth subsystem, which can be used to distinguish the sets $\{1,2,3,4\}$ and $\{1,2,4,5\}$ by the fifth party.
	
	For states $\{1,2,3,4\}$, the vector of the numbers of pairwise orthogonal relations is $(2,2,1,1,0)$. Omitting the fifth subsystem gives four quadripartite OPSs with the vector $(2,2,1,1)$, which can be perfectly distinguished by Lemma~\ref{lem:four-ops-multipartite}.
	
	For states $\{1,2,4,5\}$, the vector is $(2,1,2,1,0)$, corresponding to the quadripartite vector $(2,1,2,1)$. These four states can also be perfectly distinguished by Lemma~\ref{lem:four-ops-multipartite}.
	
	\textnormal{(6)} In case (8-58), state 5 is orthogonal to states $\{2,3\}$ on the fourth subsystem, to state 1 on the third subsystem, and to state 4 on the first subsystem. According to Lemma~\ref{lem:degree-two-one-one}, the fourth party can identify state 5. The remaining states $\{1,2,3,4\}$ have the vector $(2,2,1,0,1)$. Omitting the fourth subsystem gives four quadripartite OPSs with the vector $(2,2,1,1)$, which can be perfectly distinguished by Lemma~\ref{lem:four-ops-multipartite}.
	
	\textnormal{(7)} In case (8-67), state 2 is orthogonal to state 3 on the fifth subsystem. The fifth party can perform a local measurement to distinguish the sets $\{1,2,4,5\}$ and $\{1,3,4,5\}$.
	
	For states $\{1,2,4,5\}$, the vector of the numbers of pairwise orthogonal relations is $(2,1,2,1,0)$, which reduces to the quadripartite vector $(2,1,2,1)$.
	
	For states $\{1,3,4,5\}$, the vector is $(2,2,1,1,0)$, corresponding to the quadripartite vector $(2,2,1,1)$. By Lemma~\ref{lem:four-ops-multipartite}, both sets can be perfectly distinguished by LOCC.
	
	\textnormal{(8)} In cases (8-68)--(8-70), state 1 is orthogonal to states $\{2,3\}$ on the first subsystem, to state 4 on the second subsystem, and to state 5 on the fifth subsystem. Since states 4 and 5 are not orthogonal on the first subsystem, the first party can identify state 1 by Lemma~\ref{lem:degree-two-one-one}. The remaining four states $\{2,3,4,5\}$ have the vector $(1,1,2,2,0)$, which reduces to the quadripartite vector $(1,1,2,2)$. Their perfect distinguishability follows from Lemma~\ref{lem:four-ops-multipartite}.
	
	\textnormal{(9)} In case (8-71), state 1 is orthogonal to state 5 on the fifth subsystem, which allows the fifth party to distinguish the sets $\{1,2,3,4\}$ and $\{1,3,4,5\}$.
	
	For states $\{1,2,3,4\}$, the vector of the numbers of pairwise orthogonal relations is $(2,2,1,1,0)$. Omitting the fifth subsystem gives four quadripartite OPSs with the vector $(2,2,1,1)$.
	
	For states $\{1,3,4,5\}$, the vector is $(2,1,1,1,1)$. According to Lemma~\ref{lem:four-ops-multipartite}, both sets can be perfectly distinguished by LOCC.
	
	\textnormal{(10)} In cases (8-72) and (8-73), state 1 is orthogonal to states $\{2,3\}$ on the first subsystem, to state 5 on the second subsystem, and to state 4 on the fifth subsystem. By Lemma~\ref{lem:degree-two-one-one}, the first party can identify state 1. The remaining states $\{2,3,4,5\}$ have the vector $(1,1,2,2,0)$. Since there is no orthogonality relation on the fifth subsystem, they can be regarded as four quadripartite OPSs with the vector $(1,1,2,2)$ and can be perfectly distinguished by Lemma~\ref{lem:four-ops-multipartite}.
	
	\textnormal{(11)} In cases (8-74)--(8-79), state 1 is orthogonal to states $\{2,3\}$ on the second subsystem and to states $\{4,5\}$ on the first subsystem. Since states 4 and 5 are not orthogonal on the second subsystem, the second party can identify state 1 by Lemma~\ref{lem:degree-two-two}. The remaining four states $\{2,3,4,5\}$ have the vector $(1,0,2,2,1)$, which reduces to the quadripartite vector $(1,2,2,1)$. They can be perfectly distinguished by Lemma~\ref{lem:four-ops-multipartite}.
	
	\textnormal{(12)} In cases (8-80)--(8-102), state 1 is orthogonal to states $\{2,3\}$ on the second subsystem, to state 4 on the first subsystem, and to state 5 on the third subsystem. Since states 4 and 5 are not orthogonal on the second subsystem, Lemma~\ref{lem:degree-two-one-one} implies that the second party can identify state 1. The remaining states $\{2,3,4,5\}$ have the vector $(2,0,1,2,1)$. Omitting the second subsystem gives four quadripartite OPSs with the vector $(2,1,2,1)$, which can be perfectly distinguished by Lemma~\ref{lem:four-ops-multipartite}.
	
	\textnormal{(13)} For cases (8-103) and (8-104), state 1 is orthogonal to states $\{2,3\}$ on the second subsystem, to state 4 on the first subsystem, and to state 5 on the fifth subsystem. Since states 4 and 5 are not orthogonal on the second subsystem, the second party can identify state 1 according to Lemma~\ref{lem:degree-two-one-one}. The remaining four states $\{2,3,4,5\}$ have the vector $(2,0,2,2,0)$, which reduces to the tripartite vector $(2,2,2)$. Their orthogonality graphs correspond to cases $(1\text{-}1)$ or $(1\text{-}2)$ in Fig.~\ref{fig:four-tripartite-222}, up to relabeling of the vertices and permutation of subsystems. By Lemma~\ref{lem:tripartite-222-dist}, these four states can be perfectly distinguished by LOCC.
	
	\textnormal{(14)} In cases (8-105)--(8-112), state 1 is orthogonal to states $\{2,3\}$ on the second subsystem, to state 5 on the first subsystem, and to state 4 on the third subsystem. Since states 4 and 5 are not orthogonal on the second subsystem, the second party can identify state 1 by Lemma~\ref{lem:degree-two-one-one}. The remaining states $\{2,3,4,5\}$ have the vector $(2,0,1,2,1)$, which reduces to the quadripartite vector $(2,1,2,1)$. They can be perfectly distinguished by Lemma~\ref{lem:four-ops-multipartite}.
	
	\textnormal{(15)} In cases (8-113)--(8-120), state 1 is orthogonal to states $\{2,3\}$ on the second subsystem and to states $\{4,5\}$ on the third subsystem. Since states 4 and 5 are not orthogonal on the second subsystem, the second party can identify state 1 by Lemma~\ref{lem:degree-two-two}. The remaining states $\{2,3,4,5\}$ have the vector $(3,0,0,2,1)$. Omitting the second and third subsystems gives four tripartite OPSs with the vector $(3,2,1)$, which can be perfectly distinguished by Lemma~\ref{lem:tripartite-411-321}.
	
	\textnormal{(16)} Finally, in cases (8-121) and (8-122), state 1 is orthogonal to states $\{2,3\}$ on the second subsystem, to state 4 on the third subsystem, and to state 5 on the fourth subsystem. Since states 4 and 5 are not orthogonal on the second subsystem, the second party can identify state 1 by Lemma~\ref{lem:degree-two-one-one}. The remaining four states $\{2,3,4,5\}$ have the vector $(3,0,1,1,1)$. Omitting the second subsystem gives four quadripartite OPSs with the vector $(3,1,1,1)$, which can be perfectly distinguished by Lemma~\ref{lem:four-ops-multipartite}.
	
	Therefore, the 122 cases shown in Fig.~\ref{fig:32221-2} can be perfectly distinguished by LOCC.

	\subsection{Category $(2,2,2,2,2)$}

	For the category $(2,2,2,2,2)$, the sets of five OPSs corresponding to 13 orthogonality graphs can be perfectly distinguished by LOCC. These graphs are labeled $(9\text{-}1)$--$(9\text{-}13)$, as shown in Fig.~\ref{fig:22222}.
	
	\textnormal{(1)} In cases (9-1)--(9-3), state 1 is orthogonal to states $\{2,3\}$ on the first subsystem and to states $\{4,5\}$ on the second subsystem. Since states 4 and 5 are not orthogonal on the first subsystem, the first party can identify state 1 by Lemma~\ref{lem:degree-two-two}. The remaining four states $\{2,3,4,5\}$ have the vector $(0,0,2,2,2)$. Omitting the first and second subsystems gives four tripartite OPSs with the vector $(2,2,2)$. Their orthogonality graphs correspond to cases (1-1) or (1-2) in Fig.~\ref{fig:four-tripartite-222}. By Lemma~\ref{lem:tripartite-222-dist}, these four states can be perfectly distinguished by LOCC.
	
	\textnormal{(2)} In cases (9-4)--(9-13), state 1 is orthogonal to states $\{2,3\}$ on the first subsystem, to state 4 on the second subsystem, and to state 5 on the third subsystem. Since states 4 and 5 are not orthogonal on the first subsystem, the first party can identify state 1 according to Lemma~\ref{lem:degree-two-one-one}. The remaining states $\{2,3,4,5\}$ have the vector $(0,1,1,2,2)$, which reduces to the quadripartite vector $(1,1,2,2)$. These four states can be perfectly distinguished by Lemma~\ref{lem:four-ops-multipartite}.
	
	Therefore, the 13 cases shown in Fig.~\ref{fig:22222} can be perfectly distinguished by LOCC.
	

	\bibliographystyle{fop_reference}
	\bibliography{references_fop}

@article{Bennett1999QuantumNonlocality,
	author  = {Bennett, C. H. and DiVincenzo, D. P. and Fuchs, C. A. and Mor, T. and Rains, E. and Shor, P. W. and Smolin, J. A. and Wootters, W. K.},
	title   = {Quantum nonlocality without entanglement},
	journal = {Phys. Rev. A},
	volume  = {59},
	pages   = {1070},
	year    = {1999}
}

@article{Bennett1999UPB,
	author  = {Bennett, C. H. and DiVincenzo, D. P. and Mor, T. and Shor, P. W. and Smolin, J. A. and Terhal, B. M.},
	title   = {Unextendible product bases and bound entanglement},
	journal = {Phys. Rev. Lett.},
	volume  = {82},
	pages   = {5385},
	year    = {1999}
}

@article{Walgate2000LocalDistinguishability,
	author  = {Walgate, J. and Short, A. J. and Hardy, L. and Vedral, V.},
	title   = {Local distinguishability of multipartite orthogonal quantum states},
	journal = {Phys. Rev. Lett.},
	volume  = {85},
	pages   = {4972},
	year    = {2000}
}

@article{DiVincenzo2003Unextendible,
	author  = {DiVincenzo, D. P. and Terhal, B. M. and Thapliyal, A. V. and Leung, D. W. and Smolin, J. A.},
	title   = {Unextendible product bases, uncompletable product bases and bound entanglement},
	journal = {Commun. Math. Phys.},
	volume  = {238},
	pages   = {379},
	year    = {2003}
}

@article{Walgate2002Nonlocality,
	author  = {Walgate, J. and Hardy, L.},
	title   = {Nonlocality, asymmetry, and distinguishing bipartite states},
	journal = {Phys. Rev. Lett.},
	volume  = {89},
	pages   = {147901},
	year    = {2002}
}

@article{Zhang2014NonlocalityOPB,
	author  = {Zhang, Z. C. and Gao, F. and Tian, G. J. and Cao, T. Q. and Wen, Q. Y.},
	title   = {Nonlocality of orthogonal product basis quantum states},
	journal = {Phys. Rev. A},
	volume  = {90},
	pages   = {022313},
	year    = {2014}
}

@article{Zhang2015NonlocalityOPS,
	author  = {Zhang, Z. C. and Gao, F. and Qin, S. J. and Yang, Y. H. and Wen, Q. Y.},
	title   = {Nonlocality of orthogonal product states},
	journal = {Phys. Rev. A},
	volume  = {92},
	pages   = {012332},
	year    = {2015}
}

@article{Wang2015NonlocalityOPB,
	author  = {Wang, Y. L. and Li, M. S. and Zheng, Z. J. and Fei, S. M.},
	title   = {Nonlocality of orthogonal product-basis quantum states},
	journal = {Phys. Rev. A},
	volume  = {92},
	pages   = {032313},
	year    = {2015}
}

@article{Xu2016MultipartiteNonlocality,
	author  = {Xu, G. B. and Wen, Q. Y. and Qin, S. J. and Yang, Y. H. and Gao, F.},
	title   = {Quantum nonlocality of multipartite orthogonal product states},
	journal = {Phys. Rev. A},
	volume  = {93},
	pages   = {032341},
	year    = {2016}
}

@article{Xu2016ArbitraryBipartite,
	author  = {Xu, G. B. and Yang, Y. H. and Wen, Q. Y. and Qin, S. J. and Gao, F.},
	title   = {Locally indistinguishable orthogonal product bases in arbitrary bipartite quantum system},
	journal = {Sci. Rep.},
	volume  = {6},
	pages   = {31048},
	year    = {2016}
}

@article{Zhang2017Construction,
	author  = {Zhang, Z. C. and Zhang, K. J. and Gao, F. and Wen, S. Q. and Oh, C. H.},
	title   = {Construction of nonlocal multipartite quantum states},
	journal = {Phys. Rev. A},
	volume  = {95},
	pages   = {052344},
	year    = {2017}
}

@article{Halder2018SeveralNonlocalSets,
	author  = {Halder, S.},
	title   = {Several nonlocal sets of multipartite pure orthogonal product states},
	journal = {Phys. Rev. A},
	volume  = {98},
	pages   = {022303},
	year    = {2018}
}

@article{Jiang2020NonlocalSetsArbitrary,
	author  = {Jiang, D. H. and Xu, G. B.},
	title   = {Nonlocal sets of orthogonal product states in an arbitrary multipartite quantum system},
	journal = {Phys. Rev. A},
	volume  = {102},
	pages   = {032211},
	year    = {2020}
}

@article{Zhu2022TripartiteLessElements,
	author  = {Zhu, Y. Y. and Jiang, D. H. and Liang, X. Q. and Xu, G. B. and Yang, Y. G.},
	title   = {Nonlocal sets of orthogonal product states with the less amount of elements in tripartite quantum systems},
	journal = {Quantum Inf. Process.},
	volume  = {21},
	pages   = {252},
	year    = {2022}
}

@article{Zhen2022GeneralMultipartite,
	author  = {Zhen, X. F. and Fei, S. M. and Zuo, H. J.},
	title   = {Nonlocality without entanglement in general multipartite quantum systems},
	journal = {Phys. Rev. A},
	volume  = {106},
	pages   = {062432},
	year    = {2022}
}

@article{Zhen2025LocallyStable,
	author  = {Zhen, X. F. and Qin, S. J. and Ma, T. P. and Zuo, H. J. and Gao, F.},
	title   = {Locally stable sets without entanglement with minimum cardinality},
	journal = {Phys. Rev. A},
	volume  = {111},
	pages   = {052344},
	year    = {2025}
}

@article{Halder2019StrongQuantumNonlocality,
	author  = {Halder, S. and Banik, M. and Agrawal, S. and Bandyopadhyay, S.},
	title   = {Strong quantum nonlocality without entanglement},
	journal = {Phys. Rev. Lett.},
	volume  = {122},
	pages   = {040403},
	year    = {2019}
}

@article{Yuan2020StrongQuantumNonlocality,
	author  = {Yuan, P. and Tian, G. and Sun, X. M.},
	title   = {Strong quantum nonlocality without entanglement in multipartite quantum systems},
	journal = {Phys. Rev. A},
	volume  = {102},
	pages   = {042228},
	year    = {2020}
}

@article{Shi2020StrongWithEntanglement,
	author  = {Shi, F. and Hu, M. Y. and Chen, L. and Zhang, X.},
	title   = {Strong quantum nonlocality with entanglement},
	journal = {Phys. Rev. A},
	volume  = {102},
	pages   = {042202},
	year    = {2020}
}

@article{Wang2021GraphConnectivity,
	author  = {Wang, Y. L. and Li, M. S. and Yung, M. H.},
	title   = {Graph-connectivity based strong quantum nonlocality with genuine entanglement},
	journal = {Phys. Rev. A},
	volume  = {104},
	pages   = {012424},
	year    = {2021}
}

@article{Shi2022UPBHeterogeneous,
	author  = {Shi, F. and Li, M. S. and Chen, L. and Zhang, X.},
	title   = {Strong quantum nonlocality for unextendible product bases in heterogeneous systems},
	journal = {J. Phys. A: Math. Theor.},
	volume  = {55},
	pages   = {015305},
	year    = {2022}
}

@article{Shi2022NPartite,
	author  = {Shi, F. and Ye, Z. and Chen, L. and Zhang, X.},
	title   = {Strong quantum nonlocality in n-partite systems},
	journal = {Phys. Rev. A},
	volume  = {105},
	pages   = {022209},
	year    = {2022}
}

@article{Zhou2023EvenN,
	author  = {Zhou, H. Q. and Gao, T. and Yan, F. L.},
	title   = {Strong quantum nonlocality without entanglement in an n-partite system with even n},
	journal = {Phys. Rev. A},
	volume  = {107},
	pages   = {042214},
	year    = {2023}
}

@article{Zhu2023MinimalNonlocality,
	author  = {Zhu, Y. Y. and Jiang, D. H. and Xu, G. B. and Yang, Y. G.},
	title   = {Completable sets of orthogonal product states with minimal nonlocality},
	journal = {Physica A},
	volume  = {624},
	pages   = {128956},
	year    = {2023}
}

@article{Xu2025MinimalNonlocality,
	author  = {Xu, G. B. and Zhang, Y. Q. and Jiang, D. H.},
	title   = {Minimal nonlocality in multipartite quantum systems},
	journal = {Phys. Rev. A},
	volume  = {111},
	pages   = {022416},
	year    = {2025}
}

@article{Xu2025FourMultipartiteOPS,
	author  = {Xu, G. B. and Hao, Z. Y. and Wang, H. K. and Yang, Y. G. and Jiang, D. H.},
	title   = {Local distinguishability of four multipartite orthogonal product states},
	journal = {Phys. Rev. A},
	volume  = {112},
	pages   = {042433},
	year    = {2025}
}

@article{Xu2026FiveBipartiteTripartiteOPS,
	author  = {Xu, G. B. and Hao, Z. Y. and Wang, H. K. and Yang, Y. G. and Jiang, D. H.},
	title   = {Local distinguishability of five orthogonal product states on bipartite and tripartite quantum systems},
	journal = {Phys. Rev. A},
	volume  = {113},
	pages   = {052436},
	year    = {2026}
}

@book{BondyMurty2008GraphTheory,
	author    = {Bondy, J. A. and Murty, U. S. R.},
	title     = {Graph Theory},
	publisher = {Springer},
	address   = {London},
	year      = {2008}
}

@misc{Xu2026SixBipartiteOPS,
	author        = {Xu, G. B. and Wang, H. K. and Yang, Y. G. and Jiang, D. H.},
	title         = {Local distinguishability of six bipartite orthogonal product states},
	howpublished = {arXiv:2607.02006 [quant-ph]},
	year          = {2026},
	eprint        = {2607.02006},
	archivePrefix = {arXiv},
	primaryClass  = {quant-ph}
}

@misc{Xu2026FiveQuadripartiteOPS,
	author        = {Xu, G. B. and Zhan, J. and Yang, Y. G. and Jiang, D. H.},
	title         = {Local distinguishability of five quadripartite orthogonal product states},
	howpublished = {arXiv:2610.29622 [quant-ph]},
	year          = {2026},
	eprint        = {2610.29622},
	archivePrefix = {arXiv},
	primaryClass  = {quant-ph}
}

\end{document}